\documentstyle[11pt,epsf]{article}
\input epsf.tex
 1
 1
 1

\def\lsim{\mathrel{\raise.3ex\hbox{$<$\kern-.75em\lower1ex\hbox{$\sim$}}}}
\def\gsim{\mathrel{\raise.3ex\hbox{$>$\kern-.75em\lower1ex\hbox{$\sim$}}}}
\begin{document}
\noindent
\thispagestyle{empty}
\renewcommand{\thefootnote}{\fnsymbol{footnote}}
\begin{flushright}
{\bf CERN-TH/99-282}\\
{\bf McGill/99-29}\\
{\bf hep-ph/9909495}\\
\end{flushright}
\vspace{.5cm}
\begin{center}
  \begin{Large}\bf
Details on the ${\cal{O}}(m_e \alpha^6)$ Positronium Hyperfine
Splitting \\[2mm]
due to Single Photon Annihilation
  \end{Large}
  \vspace{1.5cm}

\begin{large}
 A.H.~Hoang$^{a}$,
 P.~Labelle$^{b}$ and
 S.M.~Zebarjad$^{c}$
\end{large}

\vspace{1.5cm}
\begin{it}
${}^a$ Theory Division, CERN,\\
   CH-1211 Geneva 23, Switzerland\\[.5cm]
${}^b$  Department of Physics, McGill University,\\
   Montr\'eal, Qu\'ebec, Canada H3A 2T8\\[.5cm]
${}^c$ Physics Department and Biruni Observatory,\\
Shiraz University, Shiraz 71454, Iran 
\end{it}

  \vspace{2.5cm}
  {\bf Abstract}\\
\vspace{0.3cm}

\noindent
\begin{minipage}{14.0cm}
\begin{small}
A detailed presentation is given of the analytic calculation of the
single-photon annihilation contributions for the positronium ground
state hyperfine splitting, to order $m_e\alpha^6$ in the framework of
non-relativistic effective theories. The current status of the theoretical
description of the positronium ground state hyperfine splitting is reviewed.
\\[3mm]
\end{small}
\end{minipage}
\end{center}
\setcounter{footnote}{0}
\renewcommand{\thefootnote}{\arabic{footnote}}
\vspace{1.2cm}
{\bf CERN-TH/99-282}\\
{\bf September 1999}
%
%
%
\newpage
\noindent

\section{Introduction}
\label{sectionintroduction}

Quantum electrodynamics is the prototype of a quantum field theory, and
its successes in describing the interactions of leptons and photons
have been spectacular. Nevertheless, continuous quantitative tests of QED,
particularly at the level of high precision, are important. The
positronium system, a two-body bound state consisting of an electron and
a positron, provides a clean testing ground of QED because the effects
of the strong and the electroweak interactions are negligible, even at
the present accuracy of experimental measurements. The existence of
positronium was predicted in 1934~\cite{Mohorovicic1} based on the
relativistic quantum theory developed by Dirac and experimentally
verified at the beginning of the 1950s~\cite{Deutsch1}. For the ground
state hyperfine splitting, the energy difference between the
$1{}^3\!S_1$ (ortho) and $1{}^1\!S_0$ (para) states, the most recent
experimental values read~\cite{Ritter1}
\begin{equation}
W 
\, = \, 203\,389.10(74)~\mbox{MHz}
\end{equation}
and~\cite{Mills1,Mills2}
\begin{equation}
W \, = \, 203\,387.5(1.6)~\mbox{MHz}
\,.
\end{equation}
They represent a precision of 3.6 and 7.9 ppm, respectively, which
makes the calculation of all ${\cal{O}}(\alpha^2)$ (NNLO) corrections
to the leading and next-to-leading order expression 
mandatory. Since the dominant contribution to the hyperfine
splitting is of order $m_e\alpha^4$, NNLO corrections correspond to
the contributions of order $m_e\alpha^6$. 
Including also the known order $m_e\alpha^7\ln^2\alpha^{-1}$
contributions~\cite{Karshenboim1,thesis}
the theoretical expression for the hyperfine splitting reads\footnote{
We use natural units, in which $\hbar = c = 1$.
The term $\propto m_e\alpha^6 \ln\alpha^{-1}$ has been
determined in Ref.~\cite{CaswellBodwin1}.
}  
\begin{equation}
W \, = \, 
m_e\,\alpha^4\,\bigg[\,
\frac{7}{12} - \frac{\alpha}{\pi}\,\bigg(\,
\frac{8}{9}+\frac{1}{2}\,\ln 2
\,\bigg)
+ \alpha^2\,\bigg(\,
\frac{5}{24}\,\ln\alpha^{-1} + K
\,\bigg) 
-\frac{7}{8 \pi}\,\alpha^3\,\ln^2\alpha^{-1}
\,\bigg]
\,,
\label{LOandNLO}
\end{equation}
where $\alpha$ is the fine-structure constant.
At order $m_e\alpha^6$ it is convenient to distinguish between four
different sorts of corrections: non-annihilation, single-, two- and
three-photon annihilation corrections. The two- and three-photon
annihilation contributions have been calculated analytically in
Refs.~\cite{Adkins1} and \cite{Adkins2}, respectively. The 
single-photon annihilation contributions have recently been determined in
Refs.~\cite{Adkins3,Hoang1}. In Ref.~\cite{Hoang1} an analytic result
has been presented and in Ref.~\cite{Adkins3} a numerical one; the two
results are in agreement. For the non-annihilation contributions, three
different results exist in the
literature~\cite{Caswell1,Pachucki1,Adkins4,Czarnecki1}, where
Refs.~\cite{Caswell1,Pachucki1,Adkins4} have presented numerical
results and
Ref.~\cite{Czarnecki1} analytical ones. The results of
Refs.~\cite{Pachucki1} and \cite{Czarnecki1} are in agreement. 
 
A modern and very economical method to calculate non-relativistic
bound state problems is based on the concept of effective field
theories. This approach was first proposed in
Ref.~\cite{Caswell1}. The effective field theoretical approach to the
positronium bound state problem is based on the existence of widely
separated scales in the positronium system. The physical effects
associated to these scales are separated by reformulating QED in terms
of an effective non-relativistic, non-renormalizable Lagrangian, where
the low scale effects correspond to an infinite set of operators and
the high scale effects are encoded in the coefficients of the
operators. It is the characteristic feature of the effective field
theoretical approach that it provides a set of systematic scaling (or
power counting) rules that allow for an easy identification of all
terms that contribute to a certain order in the bound state
calculation.
The results presented in
Refs.~\cite{Hoang1,Caswell1,Czarnecki1} have been obtained
within an effective field theory approach. 
It is the purpose of this
paper to present details of the analytical calculation of the order
$m_e\alpha^6$ single-photon annihilation contribution to the hyperfine
splitting presented recently in Ref.~\cite{Hoang1}. 

The program of this paper is as follows:
in Sec.~\ref{sectionconcept} we give an overview of the effective
field theory approach to the positronium bound state problem, and we 
explain the various steps in the calculation of the single-photon
annihilation contributions to the hyperfine splitting. 
Section~\ref{sectionregularization} contains a discussion of the
subtleties of the cutoff regularization prescription that we use in
our calculation. In Sec.~\ref{sectionmatching} we describe in detail
the calculation of the two-loop short-distance coefficient that is
needed to determine the single-photon annihilation contributions to the
hyperfine splitting at order $m_e\alpha^6$, and in
Sec.~\ref{sectionboundstate} we present the bound state calculation,
which leads to the final result. 
A generalization of the result for the single-photon annihilation
contributions to the hyperfine splitting to general radial
excitations is given in Sec.~\ref{sectionradial}. 
Section~\ref{sectiondiscussion} outlines the status of the
theoretical calculations of the hyperfine splitting, and
Sec.~\ref{sectionsummary} contains a summary.
At the end of this work we have attached an
appendix where we give a collection
of integrals that is useful for the matching calculation.

\vspace{1.5cm}
\section{The Conceptual Framework}
\label{sectionconcept}
The dynamics of a non-relativistic $e^+e^-$ pair bound together in the
positronium is governed by three widely separated scales: $m_e$,
$m_e\alpha$ and $m_e\alpha^2$.  
Because we are dealing with a Coulombic system, where the
electron/positron velocity $v$ is of order  $\alpha$ ($v\sim\alpha$),
we could equally well talk about the scales $m_e v$ and $m_e v^2$ instead
of $m_e\alpha$ and $m_e\alpha^2$.
These three scales govern different kinds of physical
processes of the positronium dynamics. The hard scale $m_e$ is
associated with $e^+e^-$ annihilation and production processes, the
dynamics of the small component and photons with virtuality of the
order of the electron mass. The soft scale $m_e v$ governs the binding
of the $e^+e^-$ pair into a bound state and directly sets the scale of
the size of the bound state wave function, the inverse Bohr
radius. The ultrasoft scale $m_e v^2$ is of the order of the binding
energy and governs low virtuality photon 
radiation processes. These processes are associated with higher Fock
states, where one has to consider the extended system $e^+e^-\gamma$
rather than only an electron--positron pair. Because the interactions 
between the $e^+e^-$ pair associated with a low virtuality photon can
arise with 
a temporal retardation, the effects caused by these higher Fock states
are called ``retardation effects''. The Lamb shift in hydrogen is the
most
famous effect of this sort. The effective field theoretical approach
uses the hierarchy of these scales ($m_e \gg m_e\alpha \gg
m_e\alpha^2$) to successively integrate out momenta of the order of
the hard and the soft scale, and, by the same means, to separate the
effects associated with them.
In this section we give a brief overview
onto the conceptual issues involved in this method following  
Refs.~\cite{Caswell1,Labelle1,Brambilla1,Beneke1}. 
It is the strength of the effective field theoretical
approach that it provides systematical momentum scaling rules
(also called power counting rules) which allow an easy identification
of all effects that have to be taken into account for a calculation
at a specific order. We apply these scaling
rules to show that retardation effects do not contribute to the
hyperfine splitting at order $m_e\alpha^6$.

NRQED is the effective field theory, which is obtained from QED after 
all hard electron/positron and photon momenta, and the respective
antiparticle poles associated with the small components have been
integrated out. The NRQED Lagrangian reads~\cite{Caswell1}
\begin{eqnarray}
\lefteqn{
{\cal{L}}_{\mbox{\tiny NRQED}} \, = \, 
\frac{1}{2}\,(\,{\mbox{\boldmath $E$}}^2-
    {\mbox{\boldmath $B$}}^2\,)
}\nonumber\\& &
+\, \psi^\dagger\,\bigg[\,
i D_t 
+ c_2\,\frac{{\mbox{\boldmath $D$}}^2}{2\,m_e} 
+ c_4\,\frac{{\mbox{\boldmath $D$}}^4}{8\,m_e^3}
+ \ldots 
\nonumber\\[2mm] & &
\hspace{8mm}
+  \frac{c_F\,e}{2\,m_e}\,{\mbox{\boldmath $\sigma$}}\cdot
    {\mbox{\boldmath $B$}}
+ \, \frac{c_D\,e}{8\,m_e^2}\,(\,{\mbox{\boldmath $D$}}\cdot 
  {\mbox{\boldmath $E$}}-{\mbox{\boldmath $E$}}\cdot 
  {\mbox{\boldmath $D$}}\,)
+ \frac{c_S\,e}{8\,m_e^2}\,i\,{\mbox{\boldmath $\sigma$}}\,
  (\,{\mbox{\boldmath $D$}}\times 
  {\mbox{\boldmath $E$}}-{\mbox{\boldmath $E$}}\times 
  {\mbox{\boldmath $D$}}\,)
+\ldots
 \,\bigg]\,\psi 
\nonumber\\[2mm] & &
+\, \chi^\dagger\,\bigg[\,
i D_t 
- c_2\,\frac{{\mbox{\boldmath $D$}}^2}{2\,m_e} 
- c_4\,\frac{{\mbox{\boldmath $D$}}^4}{8\,m_e^3}
+ \ldots 
\nonumber\\[2mm] & &
\hspace{8mm}
-  \frac{c_F\,e}{2\,m_e}\,{\mbox{\boldmath $\sigma$}}\cdot
    {\mbox{\boldmath $B$}}
+ \, \frac{c_D\,e}{8\,m_e^2}\,(\,{\mbox{\boldmath $D$}}\cdot 
  {\mbox{\boldmath $E$}}-{\mbox{\boldmath $E$}}\cdot 
  {\mbox{\boldmath $D$}}\,)
+ \frac{c_S\,e}{8\,m_e^2}\,i\,{\mbox{\boldmath $\sigma$}}\,
  (\,{\mbox{\boldmath $D$}}\times 
  {\mbox{\boldmath $E$}}-{\mbox{\boldmath $E$}}\times 
  {\mbox{\boldmath $D$}}\,)
+\ldots
 \,\bigg]\,\chi 
\nonumber\\[2mm] & &
- \,\frac{d_1\,e^2}{4\,m_e^2}\,
  (\psi^\dagger{\mbox{\boldmath $\sigma$}}\sigma_2\chi^*)\,
  (\chi^T\sigma_2{\mbox{\boldmath $\sigma$}}\psi)
+ \frac{d_2\,e^2}{3\,m_e^4}\,
  \frac{1}{2}\Big[\,
  (\psi^\dagger{\mbox{\boldmath $\sigma$}}\sigma_2\chi^*)\,
  (\chi^T\sigma_2{\mbox{\boldmath $\sigma$}}
    (-\mbox{$\frac{i}{2}$}
  {\stackrel{\leftrightarrow}{\mbox{\boldmath $D$}}})^2\psi)
  +\mbox{h.c.}\,\Big]
+\ldots\,,
\,,
\label{NRQEDLagrangian}
\end{eqnarray}
where $\psi$ and $\chi$ are the electron and positron Pauli spinors;
$D_t$ and ${\mbox{\boldmath $D$}}$ are the time and space components
of the gauge covariant derivative $D_\mu$, $E^i = F^{0 i}$ and $B^i =
\frac{1}{2}\epsilon^{i j k} F^{j k}$ are the electric and magnetic
components of the photon field strength tensor, and $e$ is the
electric charge. The short-distance
coefficients $c_2,c_4,c_F,c_D,c_S, d_1, d_2$, which encode the effects
from moments of order $m_e$, are normalized to one at
the Born level. The subscripts $F$, $D$ and $S$ stand for Fermi,
Darwin and spin-orbit. In Eq.~(\ref{NRQEDLagrangian}) only those terms
are displayed explicitly that are relevant to the calculation of the
single-photon annihilation contributions to the hyperfine splitting at
order $m_e\alpha^6$. The four-fermion operators in the last line of
Eq.~(\ref{NRQEDLagrangian}) are of particular importance because their
coefficients encode the short-distance (i.e. hard
momentum) effects of the single-photon annihilation process. 
In general, at order $m_e\alpha^6$ for the ground state hyperfine
splitting, the four-fermion operators shown in
Eq.~(\ref{NRQEDLagrangian}) would also contain short-distance  
effects from annihilation into three photons as well as non-annihilation
effects\footnote{
The effects associated with the two-photon
annihilation process would be encoded with the spin singlet operator
$(\psi^\dagger\sigma_2\chi^*)\,
  (\chi^T\sigma_2\psi)$ (see e.g. Ref.~\cite{Burgess1}).
}, but these are not considered here. The
corresponding contributions to the hyperfine splitting have been
computed elsewhere, using other techniques (see references given in
Secs.~\ref{sectionintroduction} and \ref{sectiondiscussion}). In the
following we show explicitly that for the NNLO calculation intended in
this work we need the perturbative expansion of the constant $d_1$ at
order $\alpha^2$. For $d_2$ the Born contribution is sufficient.

From the above Lagrangian, one may derive explicit Feynman rules, after 
fixing the gauge. As is well known, the most efficient gauge for
non-relativistic calculations is the Coulomb gauge. In that gauge, the
Coulomb (or longitudinal) photon (the time component of the vector
potential) has an energy (i.e. $k_0$) independent propagator, 
$\langle A_0 A^0\rangle \simeq 1/{\mbox{\boldmath $k$}^2}$. This means that the
interaction associated with the exchange of a Coulomb photon
corresponds to an instantaneous potential. The power counting of
diagrams containing instantaneous potentials is particularly simple
because an instantaneous propagator has no particle pole,
i.e. the scale of ${\mbox{\boldmath $k$}}$ is set by the average
momentum of the fermions $\simeq  mv$ ($\simeq m \alpha$ in  the bound
state). On the other hand, the transverse photon (the spatial component
of the vector potential) has an energy-dependent  propagator, of the
form  
$\langle A_i A^j\rangle\simeq (\sum \epsilon_i \epsilon_j^*)/
(k_0^2 - \mbox{\boldmath $k$}^2)$, where the $\epsilon$
are the physical (transverse) polarization vectors. In that case,
the propagator has a particle pole, and $k_0$ and 
$\mbox{\boldmath $k$}$ can be of order $m v$ and also of order $mv^2$
(with the condition that $k_0 \leq |\mbox{\boldmath $k$}|$
(\cite{Beneke1,Griesshammer1})). This has the consequence  
that NRQED diagrams containing transverse photons involve
contributions from these two scales and, therefore, do not
contribute to a unique order in $v$ (or $\alpha$ in a bound state). 
This fact can be easily illustrated in the context of old-fashioned
(or ``time-ordered") perturbation theory in which
the integration over the energy components (via the
residues) is done from the very beginning, and one only has to
integrate over the spatial momentum components. Usually, the covariant
approach is preferred over the old-fashioned perturbation theory because a
single covariant diagram contains several time-ordered configurations
(which are recovered by performing the contour integration 
over the energy components). However, the advantage is lost
in a non-relativistic application since the different time-ordered
diagrams generally scale differently and it is in fact a disadvantage
to combine them together. 
In old-fashioned perturbation theory, one finds that a diagram
containing an electron--positron pair and a transverse photon will
contain a propagator of the form (see \cite{Labelle1} for more details) 
\begin{equation}
{1 \over |\mbox{\boldmath $k$}|}\,
{1 \over \mbox{\boldmath $p$}_{ext}^2/m_e -
\mbox{\boldmath $p$}^2/m_e -
(\mbox{\boldmath $p$} - \mbox{\boldmath $k$})^2/2 m_e - 
|\mbox{\boldmath $k$}|}
\,,
\label{transverseprop}
\end{equation}
where $\mbox{\boldmath $p$}_{ext} \simeq 
\mbox{\boldmath $p$} \simeq m_e v$ are the external and loop
momenta of the fermions (we are working in the centre-of-mass frame).
From this, one can see that different contributions arise
depending on whether the scale of $|\mbox{\boldmath $k$}|$ is set by
$\mbox{\boldmath $p$} \simeq m_e v$ or by  
$\mbox{\boldmath $p$}^2/m_e \simeq m_e v^2$. The effects associated
with the latter scale are the retardation effects. The contributions
from both 
momentum regions will not contribute to the same order in $v$.
It is, however, possible to generalize NRQED is such a way that the
contributions associated to the different scales are coming from
separate diagrams. This is achieved by simply Taylor-expanding the
NRQED diagrams containing Eq.~(\ref{transverseprop}) 
around $\mbox{\boldmath $k$} \simeq m_e v$ and around 
$\mbox{\boldmath $k$} \simeq m_e v^2$ (the latter expansion is
equivalent to a multipole expansion of the vertices)~\cite{Labelle1}. 
One finds that the lowest order term of the expansion around 
$\mbox{\boldmath $k$} \simeq m_e v$ gives a contribution of order 
\begin{equation}
\int d^3\mbox{\boldmath $k$}\,
{1 \over |\mbox{\boldmath $k$}|}\,
{-1\over |\mbox{\boldmath $k$}|} \simeq {(m_e v)^3 \over (m_e v)^2}
\simeq m_e v
\,,
\end{equation}
whereas the lowest order term of the expansion around 
$\mbox{\boldmath $k$} \simeq m_e v^2$ gives 
\begin{equation}
\int d^3\mbox{\boldmath $k$}\, 
{1 \over |\mbox{\boldmath $k$}|}\,
{1 \over \mbox{\boldmath $p$}_{ext}^2/m_e- 
   \mbox{\boldmath $p$}^2/m_e + k} \simeq 
{(m_e v^2)^3 \over (m_e v^2)^2} \simeq m_e v^2.
\end{equation}
This shows that the dominant contribution from the transverse photon
exchange comes from the scale $\mbox{\boldmath $k$} \simeq m_e v$ 
and that, to leading order, the transverse photon propagator reduces to
$-1/\mbox{\boldmath $k$}^2$ (which corresponds to simply approximating
the transverse photon propagator $1/(k_0^2-\mbox{\boldmath $k$}^2)$ by 
$-1/\mbox{\boldmath $k$}^2$).
To leading order, the diagrams containing  transverse photons are
therefore also instantaneous and one recovers the simple power counting
rules valid for the exchange of a Coulomb photon.
At sub-leading order, things are more complicated, because both
expansions must be taken into account but, fortunately, the instantaneous 
approximation will be sufficient for the present calculation, as will
be shown below. 

Because the dominant contribution from the exchange of a transverse
photon between an electron-positron pair is suppressed by $v^2$
compared to the dominant contribution from a Coulomb photon exchange
(see the electron/positron--photon couplings involving the 
$\mbox{\boldmath $B$}$ field in Eq.~(\ref{NRQEDLagrangian})) 
all interactions at NNLO (i.e. up to order $v^2$ with respect to the
Coulomb exchange) can be written as a set of simple instantaneous
potentials. In momentum space representation they are given by  
\begin{eqnarray}
\tilde V_{\mbox{\tiny Coul}}
 ({\mbox{\boldmath $p$}},{\mbox{\boldmath $q$}}) & = & 
-\,\frac{4\,\pi\,\alpha}{|{\mbox{\boldmath $p$}}-
{\mbox{\boldmath $q$}}|^2+\lambda^2}
\,,
\label{Coulombpotential}
\\[2mm]
\tilde V_{\mbox{\tiny BF}}({\mbox{\boldmath $p$}},
  {\mbox{\boldmath $q$}}) & = & 
-\,\frac{4\,\pi\,\alpha}{m_e^2}\,\bigg[\,
\frac{|{\mbox{\boldmath $p$}\times \mbox{\boldmath $q$}}|^2
   + \frac{\lambda^2}{4}|{\mbox{\boldmath $p$}}+{\mbox{\boldmath $q$}}|^2} 
   {(|{\mbox{\boldmath $p$}}-{\mbox{\boldmath $q$}}|^2+\lambda^2)^2} 
- \frac{({\mbox{\boldmath $p$}}-{\mbox{\boldmath $q$}})\times 
   {\mbox{\boldmath $S_{-}$}}\cdot 
   ({\mbox{\boldmath $p$}}-{\mbox{\boldmath $q$}})\times 
   {\mbox{\boldmath $S_{+}$}}}
   {|{\mbox{\boldmath $p$}}-{\mbox{\boldmath $q$}}|^2+\lambda^2}
\nonumber\\[2mm] & & \qquad\quad
+ \,i\,\frac{3}{2}\,\frac{({\mbox{\boldmath $p$}}\times 
   {\mbox{\boldmath $q$}})\cdot({\mbox{\boldmath $S_{-}$}}+
   {\mbox{\boldmath $S_{+}$}})}
   {|{\mbox{\boldmath $p$}}-{\mbox{\boldmath $q$}}|^2+\lambda^2}
- \frac{1}{4}\,\frac{|{\mbox{\boldmath $p$}}-
   {\mbox{\boldmath $q$}}|^2}
   {|{\mbox{\boldmath $p$}}-{\mbox{\boldmath $q$}}|^2+\lambda^2}
\,\bigg]
\,,
\label{BreitFermipotential}
\\[2mm]
\tilde V_4({\mbox{\boldmath $p$}},{\mbox{\boldmath $q$}}) & = & 
\frac{2\,\pi\,\alpha}{m_e^2}\,d_1\,\bigg[\,
\frac{3}{4} + {\mbox{\boldmath $S_{-}$}}\cdot 
  {\mbox{\boldmath $S_{+}$}}
\,\bigg]
\,,
\label{SPAoperator}
\\[2mm]
\tilde V_{4\mbox{\tiny der}}({\mbox{\boldmath $p$}},
  {\mbox{\boldmath $q$}}) & = &  
-\,\frac{4\,\pi\,\alpha}{3 m_e^4}\,({\mbox{\boldmath $p$}}^2+
  {\mbox{\boldmath $q$}}^2)\,\bigg[\,
\frac{3}{4} + {\mbox{\boldmath $S_{-}$}}\cdot 
  {\mbox{\boldmath $S_{+}$}}
\,\bigg]
\,,
\label{SPAderoperator}
\\[2mm]
\delta \tilde H_{\mbox{\tiny kin}}({\mbox{\boldmath $p$}},{\mbox{\boldmath $q$}}) 
 &=&-(2\pi)^3 \delta^{(3)}(\mbox{\boldmath
  $p$}-\mbox{\boldmath $q$})\,\
{ \mbox{\boldmath $q$}^4 \over 4 m_e^3}
\,.
\label{kineticop}
\end{eqnarray}
Here, $\lambda$ is a small fictitious photon mass introduced to
regularize infrared divergences, $\mbox{\boldmath $S_\mp$}$ are the
electron/positron spin operators, and $\alpha$ is the fine structure
constant; $\tilde V_{\mbox{\tiny BF}}$ is the
Breit--Fermi potential in the Coulomb gauge, which includes the NNLO
relativistic corrections to the Coulomb potential from the
longitudinal and transverse photon exchange. 
The potentials $\tilde V_4$ and $V_{4\mbox{\tiny der}}$ come from the
four-fermion operators in Eq.~(\ref{NRQEDLagrangian}) and
account for the single-photon annihilation process at leading order
and NNLO in the non-relativistic expansion. For convenience we will
also count the NNLO kinetic energy correction in Eq.~(\ref{kineticop})
as a potential.

Using the potentials given above, it is straightforward to derive the
momentum space equation of motion for an off-shell, time-independent
$e^+e^-e^+e^-$ four-point function in the centre-of-mass frame, valid
up to NNLO: 
\begin{equation}
\bigg[\,
 \frac{\mbox{\boldmath $p$}^2}{m_e} 
\,-\,E
\,\bigg]\,
\tilde G(\mbox{\boldmath $p$},\mbox{\boldmath $q$};s) \,+\, 
\int\frac{d^3 \mbox{\boldmath $q$}^\prime}{(2\,\pi)^3}\,
\tilde V(\mbox{\boldmath $p$},\mbox{\boldmath $q$}^\prime)\,
\tilde G(\mbox{\boldmath $q$},\mbox{\boldmath $q$}^\prime;s)
\, = \, 
(2\,\pi)^3\,\delta^{(3)}(\mbox{\boldmath $p$}-\mbox{\boldmath $q$}) 
\,,
\label{NNLOSchroedinger}
\end{equation}
where
\begin{equation}
E \, \equiv \, \sqrt{s}-2\,m_e
\end{equation}
is the centre-of-mass energy relative to the electron--positron
threshold and
\begin{eqnarray}
\tilde V(\mbox{\boldmath $p$},\mbox{\boldmath $q$}) & = & 
\tilde V_{\mbox{\tiny Coul}}(\mbox{\boldmath $p$},\mbox{\boldmath $q$}) + 
\tilde V_{\mbox{\tiny BF}}(\mbox{\boldmath $p$},
  \mbox{\boldmath $q$}) + 
\tilde V_{4}(\mbox{\boldmath $p$},
  \mbox{\boldmath $q$}) + 
\tilde V_{4\,\mbox{\tiny der}}(\mbox{\boldmath $p$},
  \mbox{\boldmath $q$}) + 
\delta\tilde H_{\mbox{\tiny kin}}(\mbox{\boldmath $p$}, \mbox{\boldmath $q$})
\,.
\label{momentumspacepotentials}
\end{eqnarray}
The equation of motion~(\ref{NNLOSchroedinger}) is a relativistic
extension of the non-relativistic Schr\"odinger equation of the
Coulomb problem. Because the potentials $\tilde V_{\mbox{\tiny BF}}$,
$\tilde V_4$, $V_{4\mbox{\tiny der}}$ and
$\delta\tilde H_{\mbox{\tiny kin}}$ lead to ultra-violet
divergences, it is important to consider Eq.~(\ref{NNLOSchroedinger})
in the framework of a consistent regularization scheme. The form of
the short-distance coefficient $d_1$ depends on the choice of the
regularization scheme. We will come back to this issue in
Sec.~\ref{sectionregularization}.

One can easily establish simple power counting rules showing that
the potentials given above are all what is needed for our 
calculation~\footnote{
We set aside subtleties arising in a cutoff
regularization scheme. Those are discussed in
Sec.~\ref{sectionregularization}.
}:
after factoring out the factors of $1/m_e$ that appears explicitly in the
potentials, the only scale left in diagrams containing the potentials
above is the inverse Bohr radius 
$\langle\mbox{\boldmath $p$}\rangle\simeq m_e\alpha$.
In order
for the final result to 
have the dimensions of energy a diagram containing any of the
potentials shown above will generate one more factor of 
$\langle\mbox{\boldmath $p$}\rangle$ than
there are factors of inverse electron mass. If there are $n$ factors
of $1/m_e$, the diagram will therefore generate a factor
$\langle\mbox{\boldmath $p$}\rangle^{n+1}/m_e^n \simeq m_e 
\alpha^{n+1}$. This is one source of powers of $\alpha$. In addition,
there are sums over intermediate states. Those contain 
a factor $1/(E_{ext} - E_{int})$, which scales like 
$1/(m_e\alpha^2)$. In order to cancel this factor of $1/m_e$, the 
diagram will also generate a factor of 
$\langle\mbox{\boldmath $p$}\rangle$, which means that each
sum over intermediate state brings in another factor of
$1/\alpha$. Finally, one must multiply by the explicit factors of
$\alpha$ contained in the NRQED vertices and in the short distance
coefficients.  
As a simple illustration of the counting rules, we may consider the
Coulomb interaction. The potential contains no inverse power of mass
(so $n=0$) and one explicit factor of $\alpha$. In first order of
perturbation theory it therefore contributes to order $m_e
\alpha^2$, which is the same order as the contribution coming from the
leading order kinetic energy. Adding one more Coulomb potential
brings in an extra factor of $\alpha$ from the vertices, but this is
cancelled by the inverse power of $\alpha$ generated by the sum over
intermediate states. The Coulomb interaction must therefore be summed
up to all orders, as is well known. This argument also shows that the 
Coulomb potential is the only interaction that must be treated
exactly, as all the other potentials contain at least two powers of
inverse mass so that adding one of those potentials leads to a
contribution of order $m \alpha^4$ (or higher).

From Eq.~(\ref{NNLOSchroedinger}) we
can now derive directly the formula for the single-photon annihilation
contribution to the hyperfine splitting at order $m_e\alpha^6$,
$W^{\mbox{\tiny 1-$\gamma$ ann}}_{\mbox{\tiny NNLO}}$. Because the
para-positronium state does not contribute, owing to C invariance, one
starts with the well-known $n=1$, ${}^3S_1$ $e^+e^-$ bound state wave
function of the non-relativistic Coulomb problem and determines  
$W^{\mbox{\tiny 1-$\gamma$ ann}}_{\mbox{\tiny NNLO}}$ via
Rayleigh--Schr\"odinger time-independent perturbation theory. Because
we are interested in the single-photon annihilation
contributions only corrections with at least one insertion of 
$\tilde V_4$
or $\tilde V_{4\mbox{\tiny der}}$ have to be taken into account. 
The formula for $W^{\mbox{\tiny 1-$\gamma$ ann}}$ at order
$m_e\alpha^6$ then reads
\begin{eqnarray}
W^{\mbox{\tiny 1-$\gamma$ ann}} & = &
\langle\,1 {}^3S_1\,| \, V_4 \, |\, 1 {}^3S_1\,\rangle
+ \langle\,1 {}^3S_1\,| \, V_{4\,\mbox{\tiny der}} \, 
      |\, 1 {}^3S_1\,\rangle
\nonumber\\[3mm] & &
+ \langle\,1 {}^3S_1\,| \, V_4 \, 
  \sum\hspace{-7.5mm}\int\limits_{l\ne 1{}^3S_1}\,
  \frac{|\,l\,\rangle\,\langle \,l\,|}{E_0-E_l} \, V_4  
  \, |\, 1 {}^3S_1\,\rangle
\nonumber\\ & &
+\,\bigg[\,
\langle\, 1 {}^3S_1\,| \,
V_4 \, \sum\hspace{-7.5mm}\int\limits_{l\ne 1{}^3S_1}\,
\frac{|\,l\,\rangle\,\langle \,l\,|}{E_0-E_l} \,
(V_{\mbox{\tiny BF}}+\delta H_{\mbox{\tiny kin}})
\, |\, 1 {}^3S_1 \,\rangle + \mbox{h.c.}
\,\bigg]\,
+\ldots
\,,
\label{Wfullformula}
\end{eqnarray} 
where $|\,l\,\rangle$ represent normalized (bound state
and continuum) eigenstates to the Coulomb Schr\"odinger
equation with the eigenvalues $E_l$; $|\,1{}^3S_1\,\rangle$ and
$E_0=-m_e\alpha^2/4$
denote the state and binding energy of the $n=1$, ${}^3S_1$ Coulomb
bound state. 
Using the counting rules developed above it is easy to show that 
Eq.~(\ref{Wfullformula}) is all we need to determine the ground state
hyperfine splitting to order $m_e\alpha^6$: 
the four-fermion operator $V_4$ contains two powers of inverse
mass and one explicit factor of $\alpha$ (with the Born level
value for the coefficient $d_1$, see Eq.~(\ref{SPAoperator})).
 The  contribution of 
this interaction is therefore of order $m_e \alpha^4$. In order to
obtain the ${\cal O}(m_e\alpha^6)$ contribution that we are looking
for, we therefore need to match the coefficient $d_1$ to two loops, as
mentioned above. The operator $V_{\mbox{\tiny 4der}}$, on the other
hand, contains 
four powers of inverse mass and  therefore contributes already to
order $m_e \alpha^6$ with the Born level coefficient given in
Eq.~(\ref{SPAderoperator}). It is easy to verify that the terms
evaluated in second order of perturbation theory also contribute to
this order if one uses the Born level coefficients in all the
potentials. Consider for example the term with two insertions of the 
potential $V_4$. Since there are four explicit powers of inverse 
mass, two explicit factors of $\alpha$ (with $d_1$ set to 1), and
one sum over intermediate states, the final contribution is of order 
$m_e \alpha^{5+2-1} = m_e \alpha^6$.
The Breit potential obviously contributes to the same order. The
operator $\delta H_{\mbox{\tiny kin}}$ does not contain any factor of
$\alpha$, but 
it contains one more power of inverse mass and therefore also
contributes to order $m_e\alpha^6$. All other potentials built from
the NRQED Feynman rules have higher powers of inverse mass and will
therefore be suppressed. We note again that the Breit--Fermi potential
$ V_{\mbox{\tiny BF}}$ contains contributions arising from the
exchange of Coulomb photons and of transverse photons in the
instantaneous approximation (i.e. without any $k_0$-dependence in the
propagator). Since, as we have shown before, the latter contribute
already to order $m_e\alpha^6$, we do not need to consider any sub-leading
terms coming from the expansions around $\mbox{\boldmath $k$} \simeq
m_e v$. Terms from the expansion around $\mbox{\boldmath $k$} \simeq
m_e v^2$ do not need to be considered at all. The instantaneous
approximation for the transverse photons is therefore sufficient for
the present calculation.  

From the above discussion, it is  clear that the calculation of
 $W^{\mbox{\tiny 1-$\gamma$ ann}}_{\mbox{\tiny NNLO}}$ proceeds
in two basic steps. 
\begin{itemize}
\item[1.]{\it Matching calculation} -- Calculation of the
${\cal{O}}(\alpha)$ and ${\cal{O}}(\alpha^2)$ contributions to the
constant $d_1$ by matching the  QED amplitudes for the
elastic s-channel scattering of free and on-shell electrons and
positrons via a single photon, $e^+e^-\to\gamma\to e^+e^-$, close to
threshold up to two loops and to NNLO in the velocity of the electrons
and positrons in the centre-of-mass frame. This is possible because
the short-distance effects encoded in $d_1$ do not depend on the
kinematic situation to which the NRQED Lagrangian is applied.
\item[2.] {\it Bound state calculation} -- Calculation of
formula on the RHS of Eq.~(\ref{Wfullformula}).
\end{itemize} 
The details of the calculations involved in steps 1 and 2 are
presented in Secs.~\ref{sectionmatching} and \ref{sectionboundstate},
respectively. 

To conclude this section we would also like to briefly mention a 
formal way to establish the multipole expansion and the counting rules
presented above. This is achieved by integrating out NRQED
electron/positron and photon momenta of order $m_e\alpha$. The
resulting effective theory has been called ``potential NRQED''
(PNRQED)~\cite{Brambilla1}. The basic ingredient to construct PNRQED is
to identify the relevant momentum regions of the electron/positron and
photon field in the NRQED Lagrangian~(\ref{NRQEDLagrangian}). These
momentum regions have been found in Ref.~\cite{Beneke1}. Because
NRQED is not Lorentz-covariant, the time and spatial components of the 
momenta are independent, which means that the time and spatial
components can have a different scaling behaviour. The relevant
momentum regions are ``soft''\footnote{
The soft momentum regime has not been taken into account in the
arguments employed in Ref.~\cite{Labelle1}. However, this does not affect
any conclusions concerning the ground state hyperfine splitting at
order $m_e\alpha^6$. 
}
($k^0\sim m_e v$, $\mbox{\boldmath $k$}\sim m_e v$),
``potential'' 
($k^0\sim m_e v^2$, $\mbox{\boldmath $k$}\sim m_e v$) and
``ultrasoft'' 
($k^0\sim m_e v^2$, $\mbox{\boldmath $k$}\sim m_e v^2$). 
It can be shown that electron, positrons and photons can have soft and
potential momenta, but that only photons can have ultrasoft momenta. A
momentum region with $k^0\sim m_e v$, $\mbox{\boldmath $k$}\sim m_e
v^2$ does not exist. PNRQED is constructed by integrating out ``soft''
electrons/positrons and photons and ``potential'' photons. In
addition, the ``potential'' photon momenta have to be expanded in
terms of their time component, because the latter scales with an
additional power of $\alpha$ with respect to the spatial components. The
exchange of ``potential'' photons between the electron and the
positron then leads to spatially non-local, but temporally
instantaneous, four-fermion operators that represent an instantaneous
coupling of an electron--positron pair separated by a distance of order
the inverse Bohr radius $\sim m_e\alpha$. The coefficients of these
operators are a 
generalization of the notion of an instantaneous potential.
Generically the PNRQED Lagrangian has the form 
\begin{eqnarray}
{\cal{L}}_{\mbox{\tiny PNRQED}} & = &
\tilde{\cal{ L}}_{\mbox{\tiny NRQED}} \, + \,
\int d^3 \mbox{\boldmath $r$} \Big(\psi^\dagger \psi\Big)
  (\mbox{\boldmath $r$})\,V(\mbox{\boldmath $r$})\,
\Big(\chi^\dagger \chi\Big)(0)
+\ldots
\,,
\label{PNRQEDLagrangian}
\end{eqnarray}
where the tilde above ${\cal{ L}}_{\mbox{\tiny NRQED}}$ on the RHS of
Eq.~(\ref{PNRQEDLagrangian}) indicates that the corresponding
operators only describe potential electrons/positrons and ultrasoft
photonic degrees of freedom and that an expansion in momentum
components $\sim m_e\alpha^2$ is understood. To NNLO, the contributions
to $V$ are just given in
Eqs.~(\ref{Coulombpotential}) to (\ref{SPAderoperator}).
Using the scaling of 
``potential'' electron/positron momenta, we see that the Coulomb potential scales
like $m_e\alpha^2$, i.e. it is of the same order as the
electron/positron kinetic energy. Thus, the Coulomb
potential has to be treated exactly rather than perturbatively. From
the PNRQED Lagrangian it is straightforward to derive the momentum
space equation of motion of an off-shell, time-independent
$(e^+e^-)(e^+e^-)$ four-point function in the centre-of-mass frame
valid up to order $\alpha^4$, Eq.~(\ref{NNLOSchroedinger}).
Using the momentum scaling rules of PNRQED one can 
show that retardation effects cannot contribute to \
$W^{\mbox{\tiny 1-$\gamma$ ann}}$ at order $m_e\alpha^6$.  
Retardation effects are caused by the ultrasoft photons, because their
low virtuality propagation can develop a pole for the momenta
available in the positronium system. Choosing again the Coulomb gauge
for our argumentation, where the time component of the Coulomb photon
vanishes, only the transverse photon needs to be considered as 
ultrasoft\footnote{ 
The argument is true in any gauge after
gauge cancellations. The argumentation is, however, most transparent
in the Coulomb gauge.
}.
Thus the emission and subsequent absorption of an ultrasoft photon between the
electron--positron pair are already suppressed by $v^2\sim\alpha^2$ with
respect to the Coulomb interaction owing to the coupling of transverse
photon to electrons/positrons. To see that an additional power of
$\alpha$ arises from the corresponding loop integration over the
ultrasoft photon momentum, let us compare the scaling of the product of
the integration measure and the photon propagator in the potential and
the ultrasoft momentum regime. In the ultrasoft case the product of
the integration measure $d^4 k$ and the 
photon propagator $1/k^2$ counts as $\alpha^8\times
\alpha^{-4}=\alpha^4$, whereas in the potential case the result reads
$\alpha^5\times \alpha^{-2}=\alpha^3$. Thus the exchange of an
ultrasoft photon is suppressed by an addition power of $\alpha$ with
respect to the effects of the Breit--Fermi
potential~(\ref{BreitFermipotential}). In other words, retardation 
effects cannot contribute to $W^{\mbox{\tiny 1-$\gamma$ ann}}$ at order
$m_e\alpha^6$. We would like to note that PNRQED is designed as a
complete field theory capable of describing the dynamics of a bound
electron--positron pair and ultrasoft photons. Although useful for
establishing consistent counting rules, its full strength only
develops if one explicitly 
considers the dynamics of ultrasoft photons. For cases where the
instantaneous approximation is sufficient -- such as the ground state
hyperfine splitting at order $m_e\alpha^6$ -- the introduction of
PNRQED is not essential.
  
\vspace{1.5cm}
\section{The Regularization Scheme}
\label{sectionregularization}

All equations in the previous section have to be considered within the
framework of a consistent UV regularization scheme. In general, the
form of the short-distance coefficients of the NRQED\footnote{
In what follows, when using the notion ``NRQED'', we actually
mean the generalized NRQED or PNRQED, as discussed in
Sec.~\ref{sectionconcept}. 
} 
operators
depends on the choice of the regularization scheme. In this work we
use a cutoff prescription to regularize the UV divergences, where the
cutoff $\Lambda$ is considered much larger than $m_e\alpha$. The infrared
divergences, which 
arise in the intermediate steps of the matching calculation to
determine the higher order contributions to $d_1$, are regularized by a
small fictitious photon mass $\lambda$, see 
Eqs.~(\ref{Coulombpotential}) and (\ref{BreitFermipotential}). The use
of a cutoff regularization involves a number of subtleties that shall
be briefly discussed in this section.

\begin{figure}[t!] 
\begin{center}
\leavevmode
\epsfxsize=6cm
\leavevmode
\epsffile[240 420 440 460]{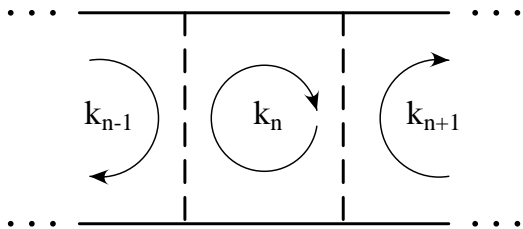}
\mbox{\hspace{6cm}}
%
%
\vskip  2.8cm
 \caption{\label{figrouting}
Routing convention for loop momenta in ladder diagrams.
Half of the external centre-of-mass energy is flowing
through each of the electron and positron lines.
}
 \end{center}
\end{figure}

It is well known that the use of a cutoff regularization scheme leads
to terms that violate gauge invariance and Ward identities. These
effects, however, are generated at the cutoff and are, therefore,
cancelled by corresponding terms with a different sign in the
short-distance coefficients of the NRQED operators. Thus, gauge
invariance and Ward identities are restored to the order at which the
matching calculation has been carried out. Another subtlety is that a
cutoff scheme is only well defined after a specific momentum routing
convention is adopted for loop diagrams in the effective field
theory. It is natural to choose the routing convention employed in the
equation of motion~(\ref{NNLOSchroedinger}). For clarity we have
illustrated this convention in Fig.~\ref{figrouting}. Because we only
need to consider interactions in our calculation that are
instantaneous in time, only ladder-type diagrams have to be taken into
account.

A very important feature of a cutoff regularization scheme is that it
inevitably leads to power-counting-breaking effects. This means that
NRQED operators can lead to effects that are below the order
indicated by the momentum scaling rules described in the previous
section. Examples of this feature will 
be visible in the matching, and the bound state calculations presented
in the next two sections. These power-counting-breaking effects are a
consequence of the fact that a cutoff regularization does not suppress
divergences of scaleless integrals (as do analytic regularization
schemes like  $\overline{\mbox{MS}}$). To illustrate the problem
let us consider the third term on the RHS
of Eq.~(\ref{Wfullformula}). This term contains the 
contribution to $W^{\mbox{\tiny 1-$\gamma$ ann}}$ coming from two
insertions of the four-fermion operator $V_4$. According to the
momentum scaling rules described in the previous section, it can only 
contribute at order $m_e\alpha^6$. However, the bound state diagram
with two $V_4$ operators is linearly divergent and, using the momentum
cutoff $\Lambda$, the result has, for dimensional reasons, the form  
$A m_e \alpha^6  + B \Lambda  \alpha^5$ where $A$ and $B$ are
finite constants (modulo logarithmic terms). If one counts
the cutoff to be of the order of $m_e$, then the 
diagram would contribute to both orders $m_e \alpha^5$ and
$m_e \alpha^6$. Even worse, by including sufficiently
high order operators (in the $p/m_e$ expansion), one can easily
convince oneself that an infinite number of operators, having much
higher dimension than indicated by the counting rules, would contribute
to any given order in $\alpha$, starting at order $m_e \alpha^5$. 
However, contributions coming from those higher-dimension operators
can only arise in the form of explicit cutoff-dependent terms and not as
constants. As for the effects that
violate gauge invariance and Ward identities, all terms depending on
the cutoff are cancelled in the combination of the bound state
integrations and the short-distance coefficient. In our perturbative
calculation we can therefore simply ignore that the scaling violating
terms coming from operators with dimensions higher than indicated by
the counting rules exist. For our calculation this means that we only
have to calculate the terms presented on the RHS of
Eq.~(\ref{Wfullformula}), excluding the higher order corrections of
$d_1$ in the third and fourth terms.  

Finally, we would like to mention that we implement our cutoff
regularization scheme in such a way that only divergent integrations
are actually cut off. This choice simplifies the
calculations, because we can use the known analytic solutions of the
non-relativistic Coulomb problem for the $1{}^3S_1$ wave function and
the Green function in our perturbative calculation. The fact that we
use a specific routing convention ensures that this does not lead to
inconsistencies. In addition, we impose the cutoff only on the spatial
components of the loop momenta.

\vspace{1.5cm}
\section{The Matching Calculation}
\label{sectionmatching}
The single-photon annihilation contributions of the short-distance
coefficient of the operator 
$(\psi^\dagger{\mbox{\boldmath $\sigma$}}\sigma_2\chi^*)\,
  (\chi^T\sigma_2{\mbox{\boldmath $\sigma$}}\psi)$ are obtained by
matching the 
amplitude for elastic s-channel scattering of an $e^+e^-$ pair via a
virtual photon in full QED in the kinematical regime close to threshold
to the same amplitude determined in the non-relativistic effective
theory NRQED. To determine the short-distance coefficient $d_1$ to
order $\alpha^2$ we have to carry out the matching at the two-loop
level, including all effects up to NNLO in the non-relativistic
expansion. Because we regulate infrared divergences in the effective
theory using a small fictitious photon mass, we have to do the same in
the full QED calculation. 

\begin{figure}[t] 
\begin{center}
\leavevmode
\epsfxsize=1.8cm
\epsffile[220 410 420 540]{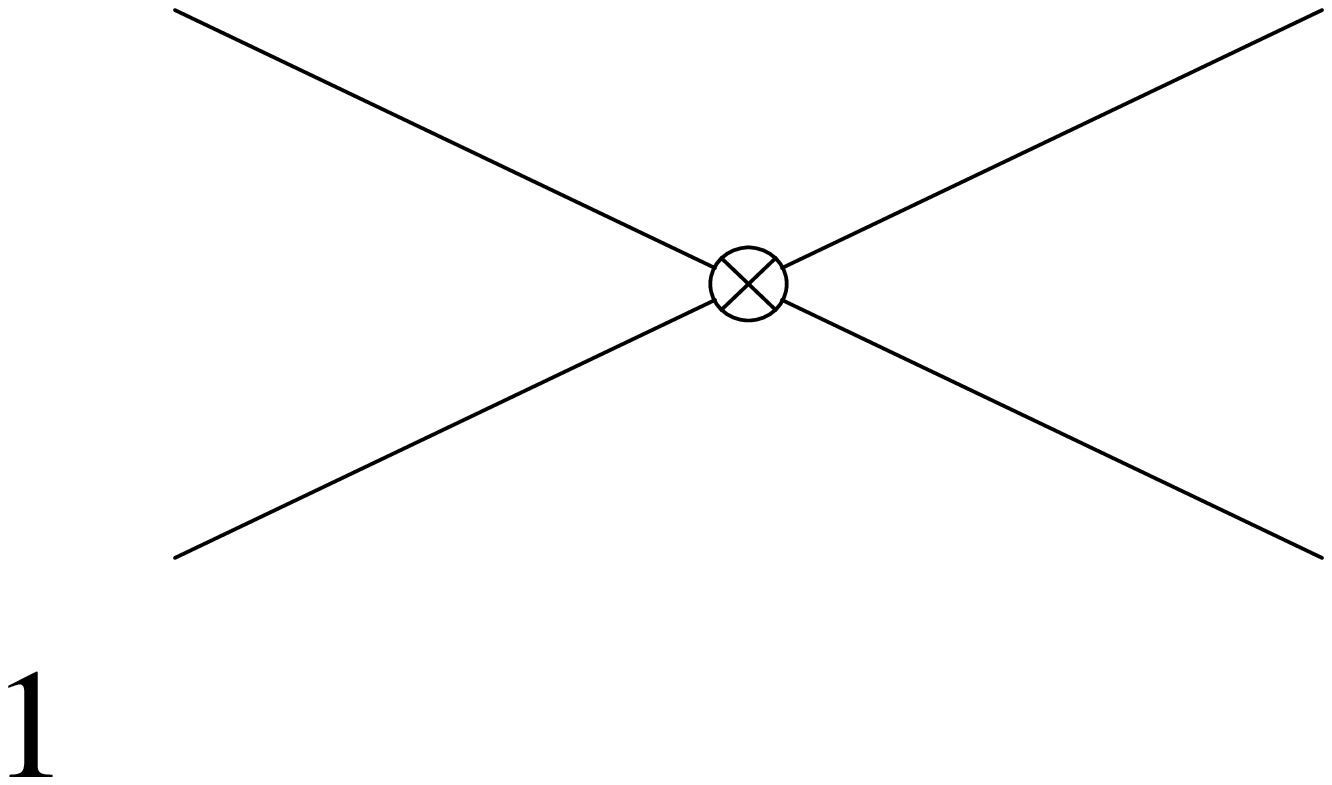}
\hspace{2.2cm}
\leavevmode
\epsfxsize=1.8cm
\epsffile[220 410 420 540]{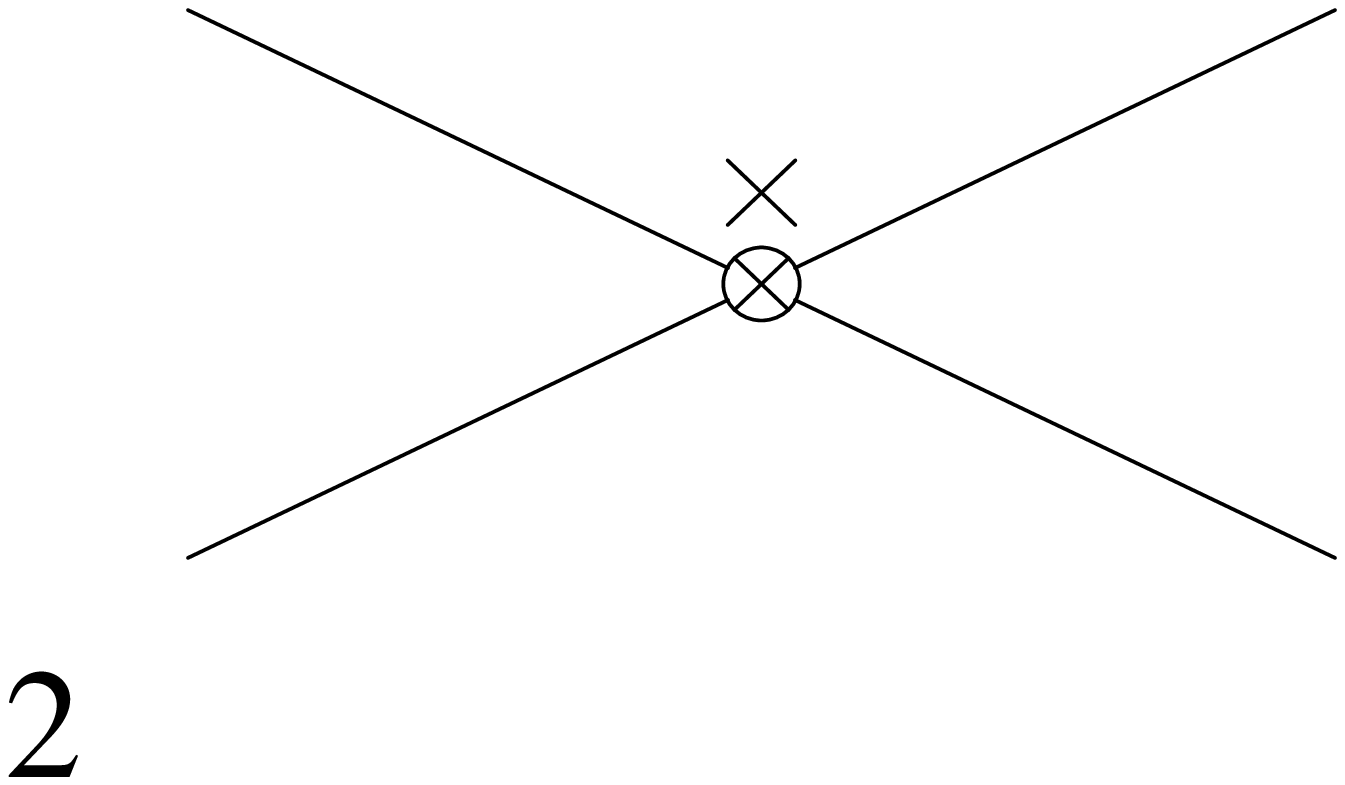}
\vskip  1.5cm
 \caption{\label{figPNRQEDborn} 
Graphical representation of the NRQED single-photon annihilation
scattering diagrams at the Born level and at NNLO in the
non-relativistic expansion.
}
 \end{center}
\end{figure}
\begin{figure}[t] 
\begin{center}
\leavevmode
\epsfxsize=1.8cm
\epsffile[220 410 420 540]{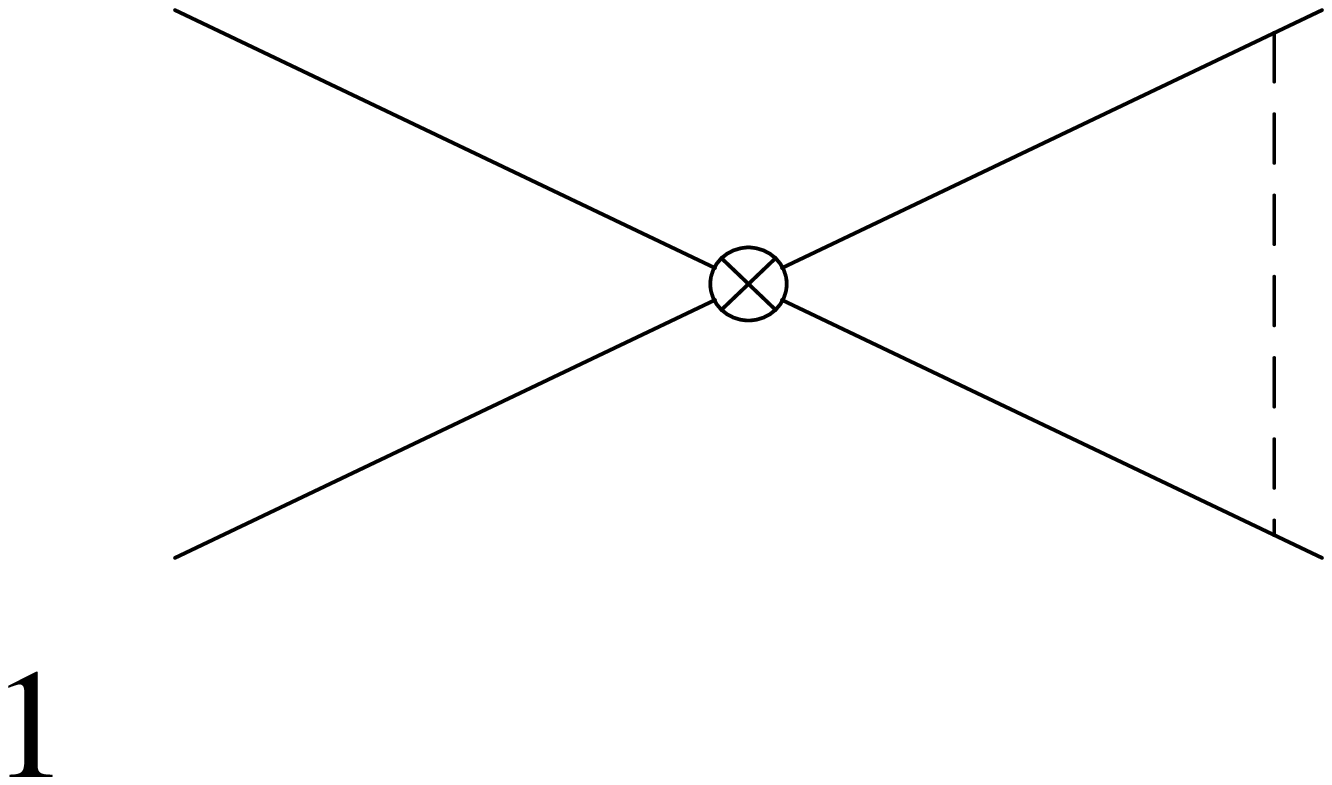}
\hspace{2.2cm}
\leavevmode
\epsfxsize=1.8cm
\epsffile[220 410 420 540]{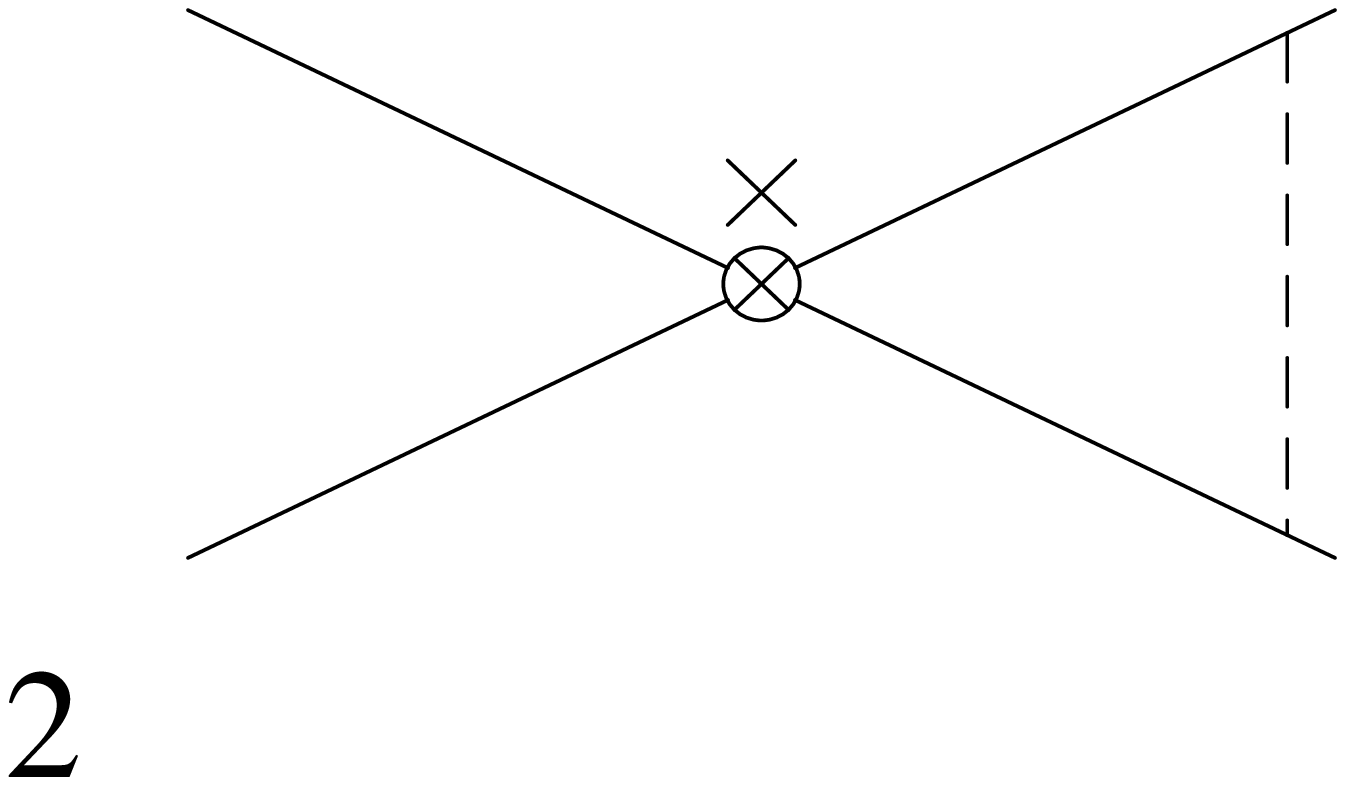}
\hspace{2.2cm}
\leavevmode
\epsfxsize=1.8cm
\epsffile[220 410 420 540]{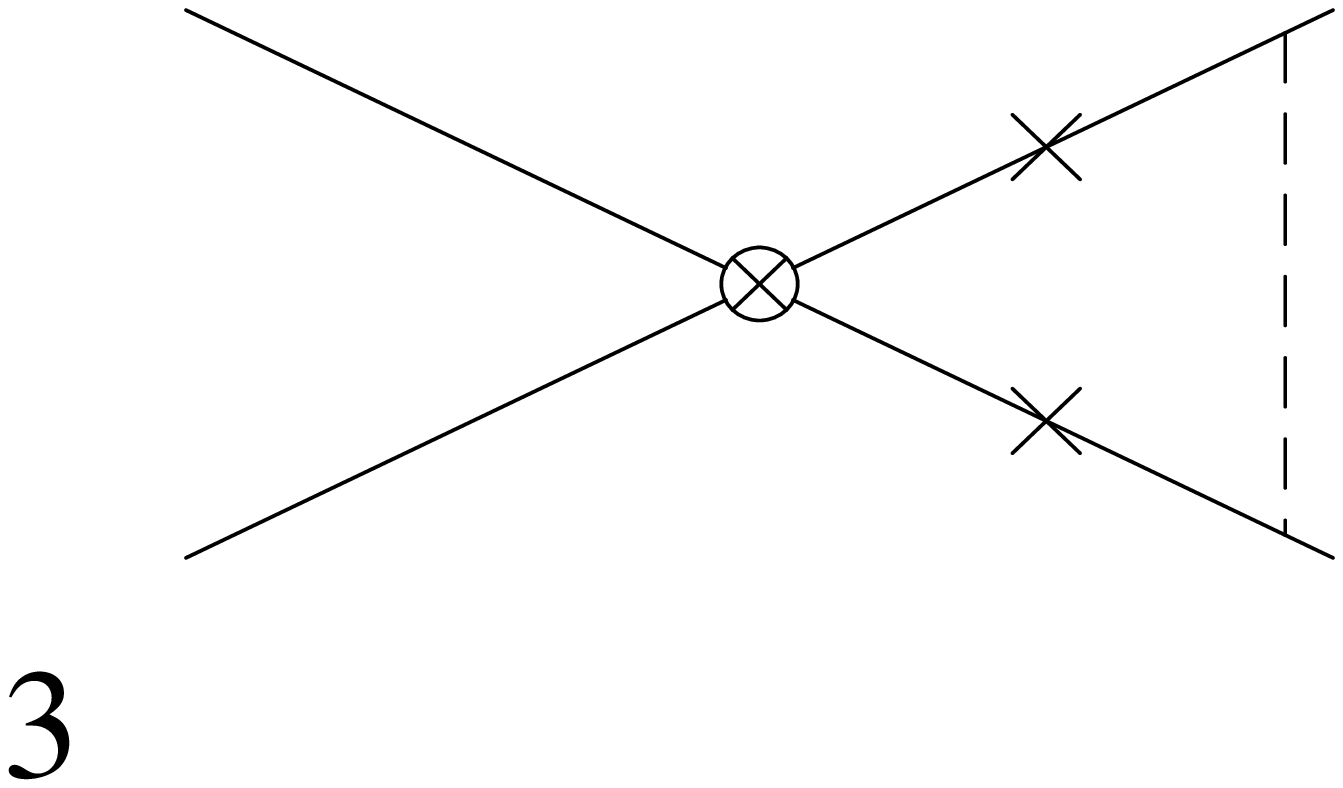}
\\[1.3cm]
\leavevmode
\epsfxsize=1.8cm
\epsffile[220 410 420 540]{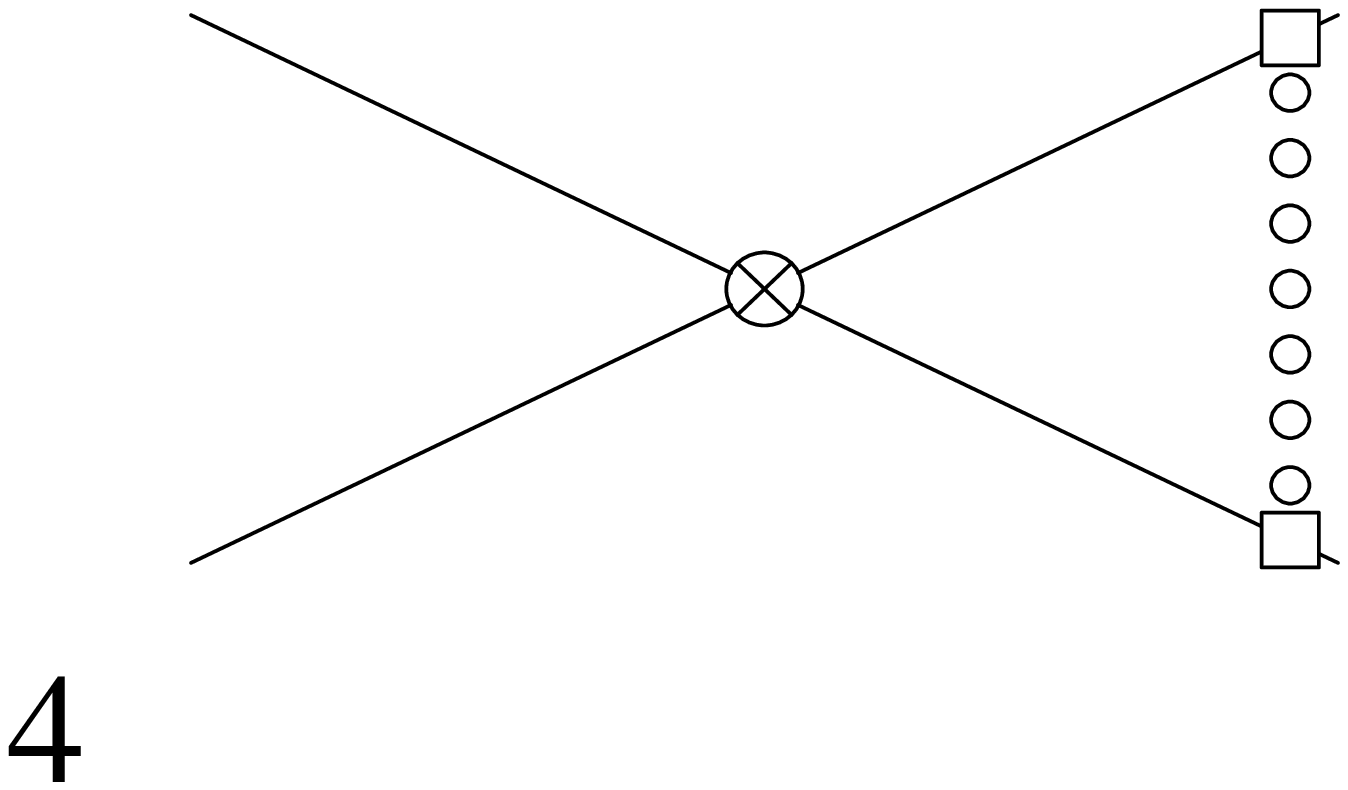}
\hspace{2.2cm}
\leavevmode
\epsfxsize=1.8cm
\epsffile[220 410 420 540]{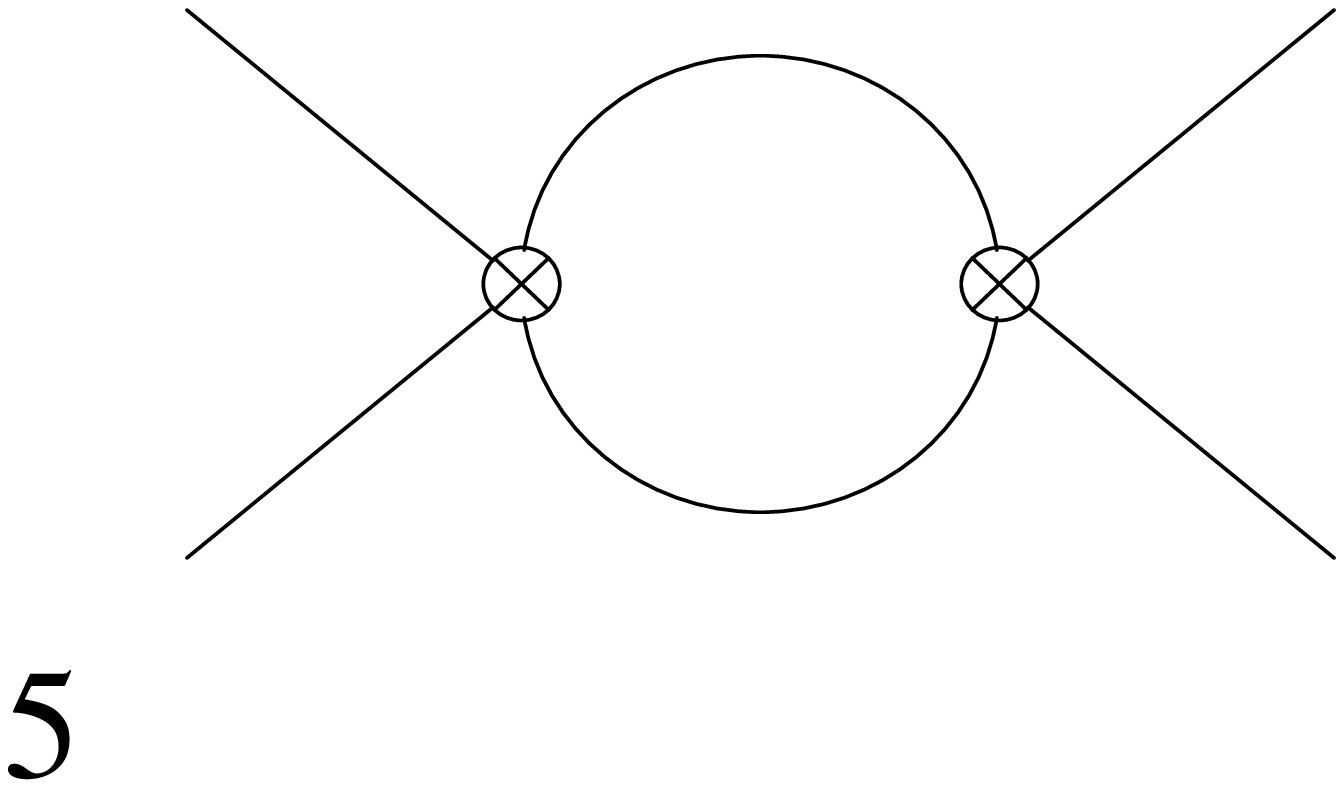}
\vskip  1.5cm
 \caption{\label{figPNRQED1loop} 
Graphical representation of the NRQED single-photon annihilation
scattering diagrams at the one-loop level and at NNLO in the
non-relativistic expansion.
}
 \end{center}
\end{figure}
\begin{figure}[t] 
\begin{center}
\leavevmode
\epsfxsize=1.8cm
\epsffile[220 410 420 540]{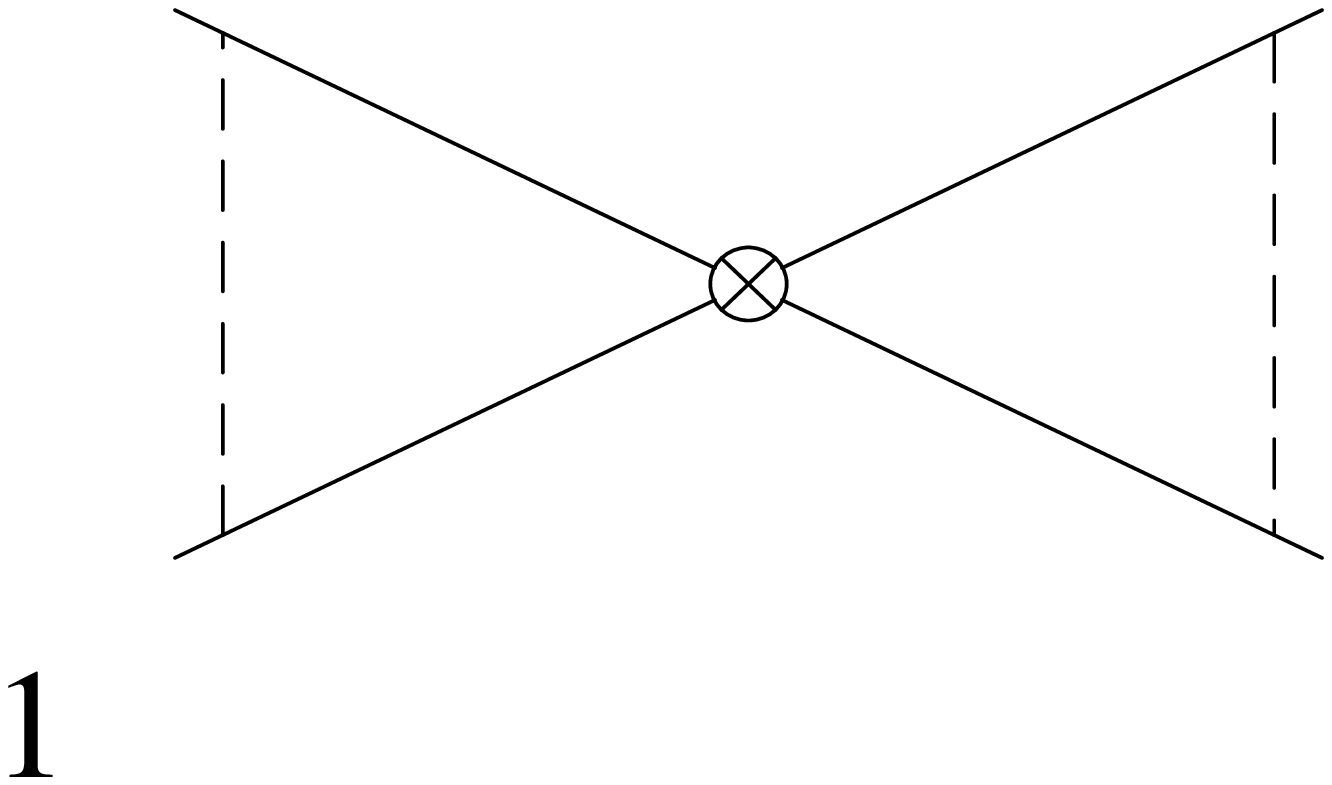}
\hspace{2.2cm}
\leavevmode
\epsfxsize=1.8cm
\epsffile[220 410 420 540]{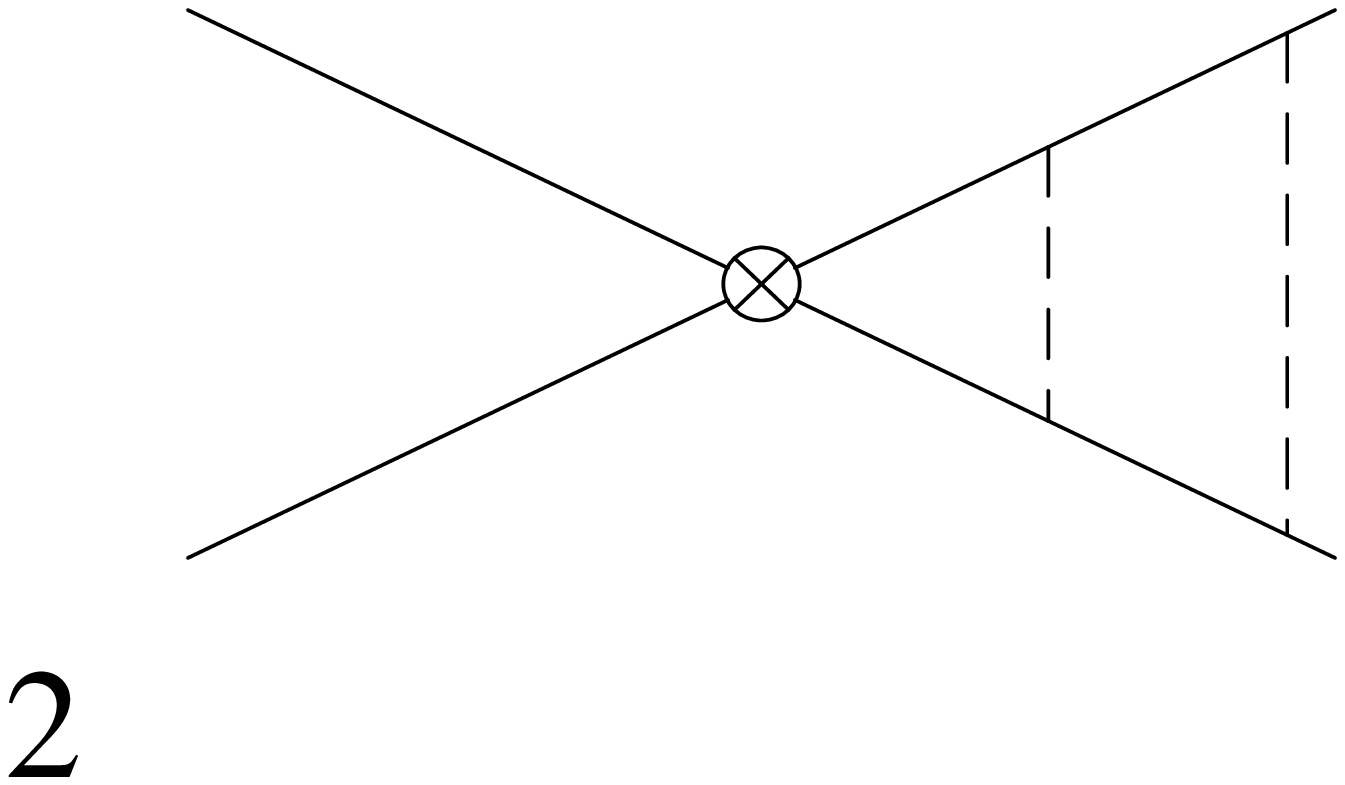}
\hspace{2.2cm}
\leavevmode
\epsfxsize=1.8cm
\epsffile[220 410 420 540]{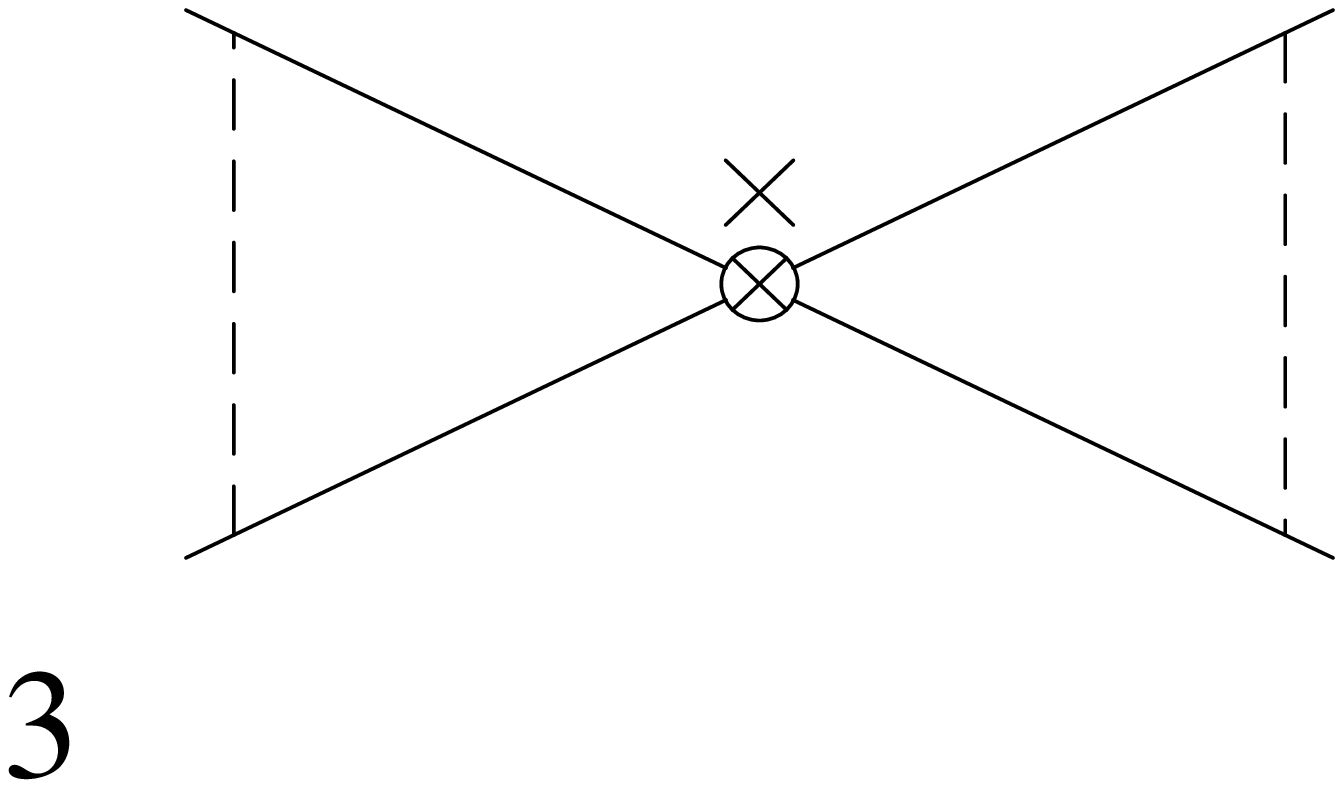}
\\[1.3cm]
\leavevmode
\epsfxsize=1.8cm
\epsffile[220 410 420 540]{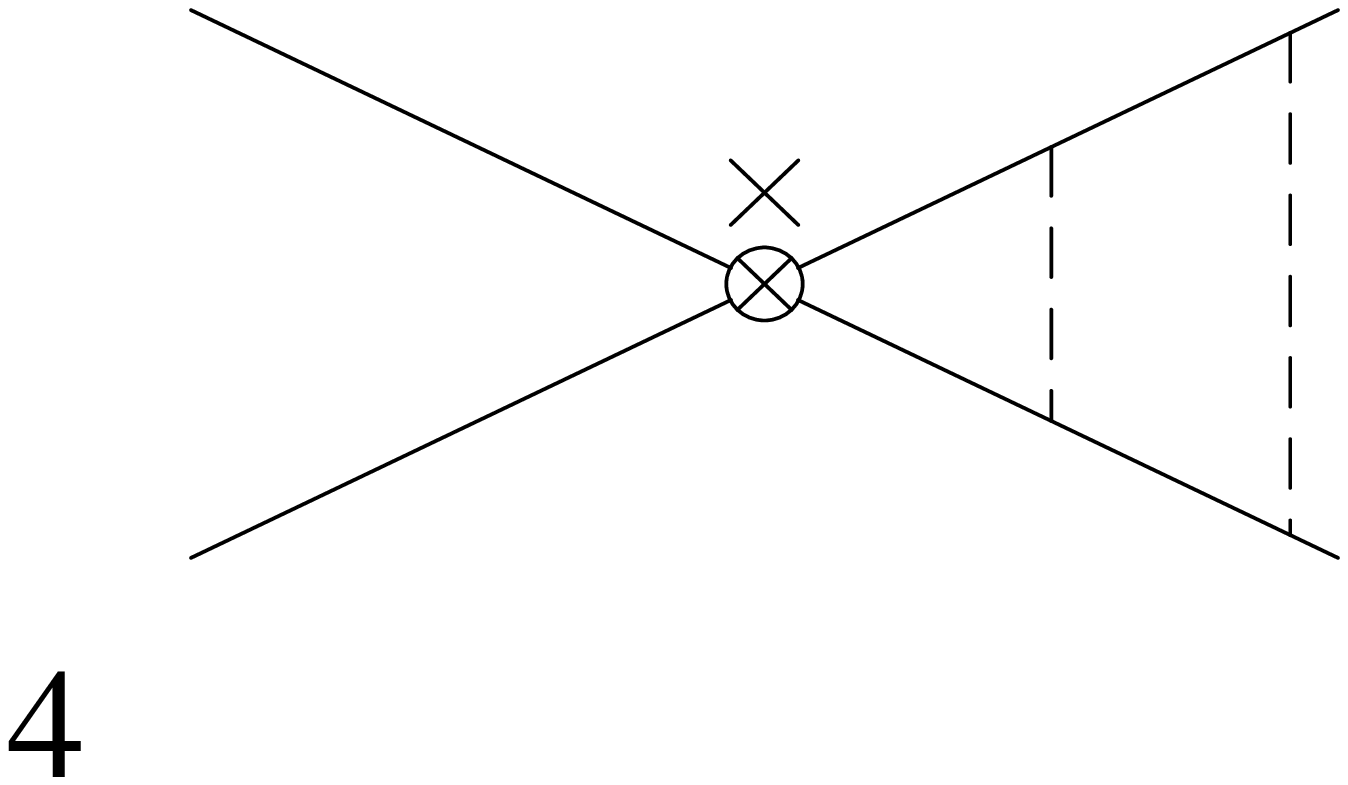}
\hspace{2.2cm}
\leavevmode
\epsfxsize=1.8cm
\epsffile[220 410 420 540]{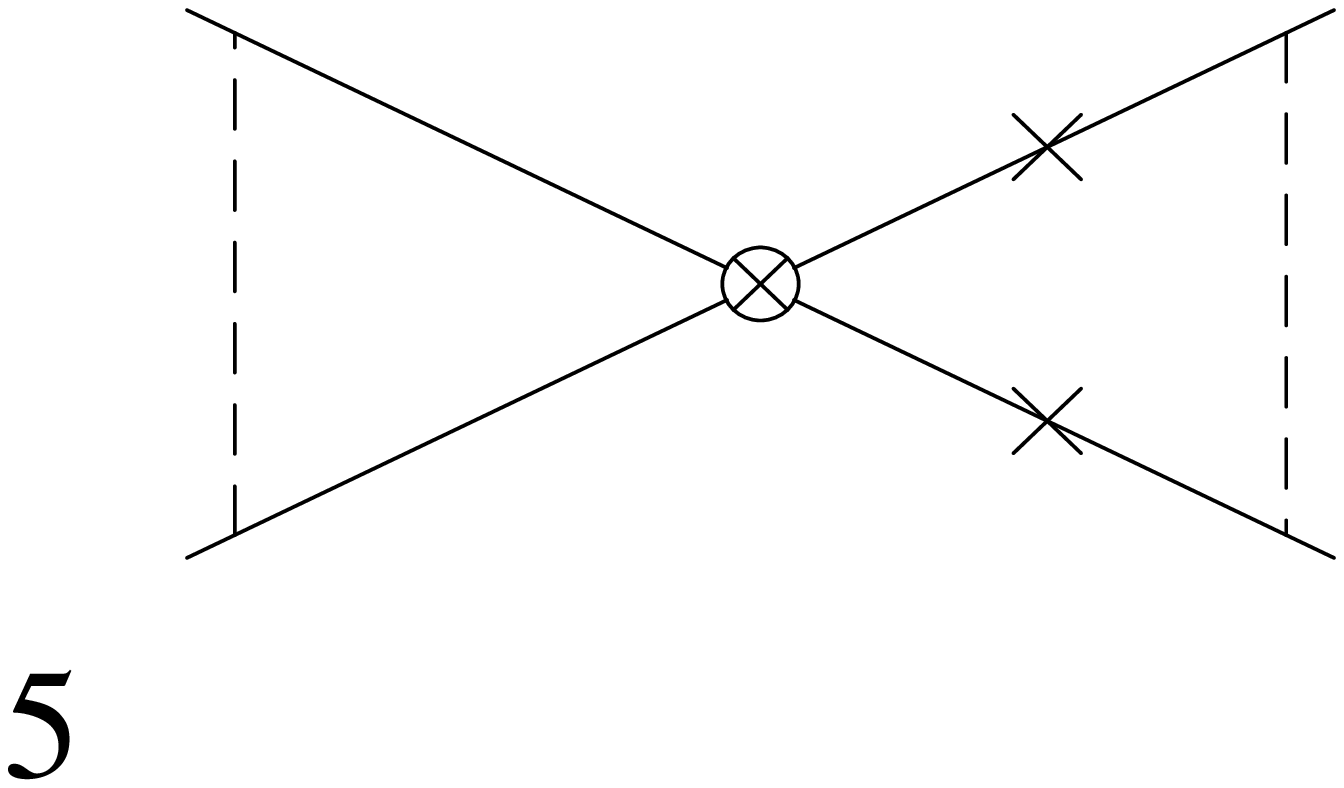}
\hspace{2.2cm}
\leavevmode
\epsfxsize=1.8cm
\epsffile[220 410 420 540]{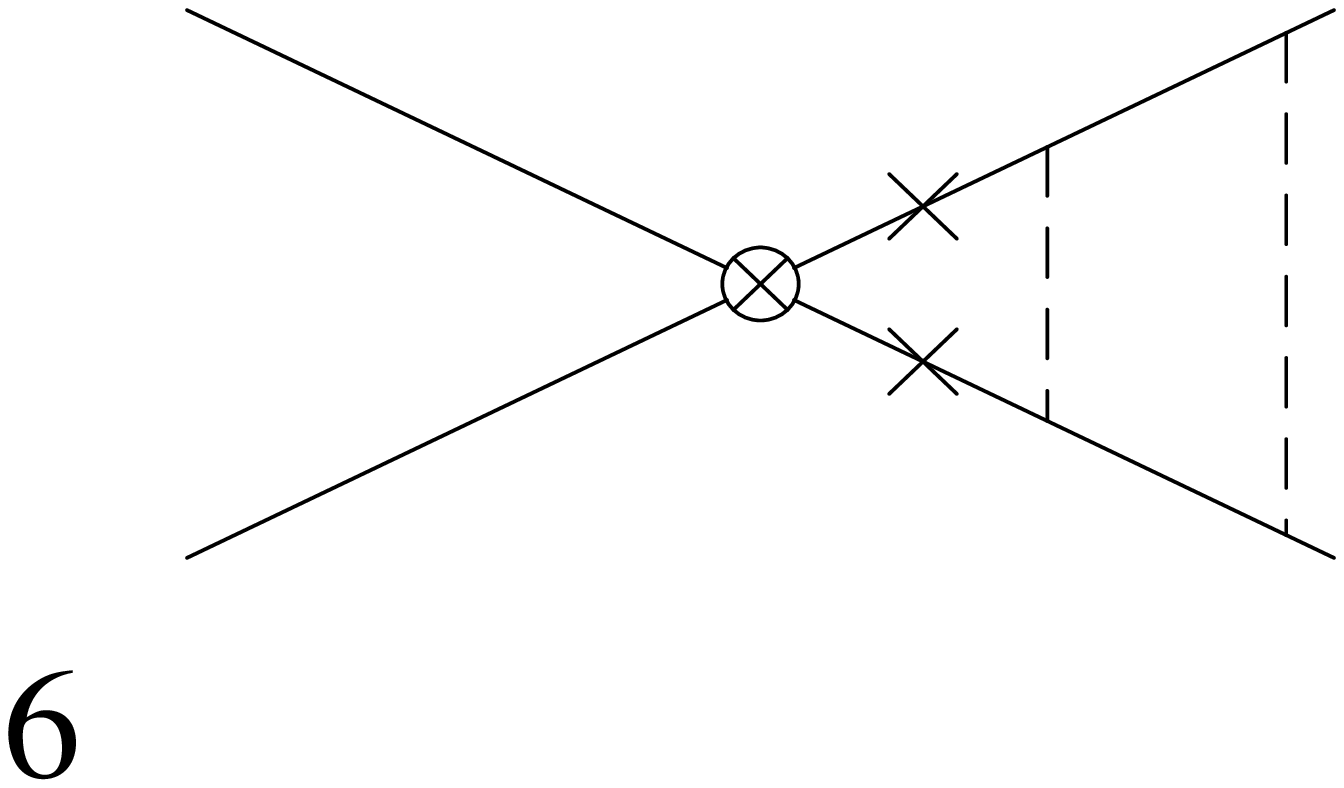}
\\[1.3cm]
\leavevmode
\epsfxsize=1.8cm
\epsffile[220 410 420 540]{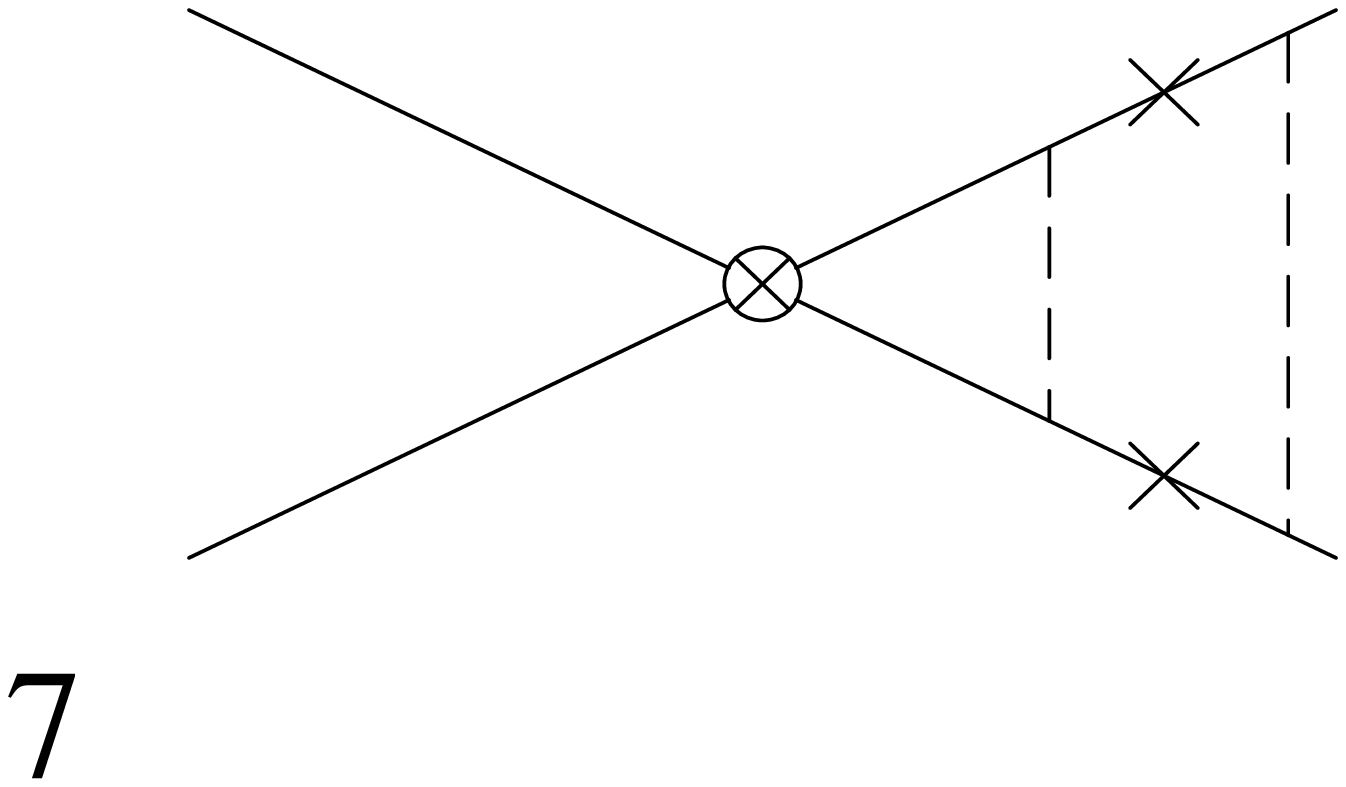}
\hspace{2.2cm}
\leavevmode
\epsfxsize=1.8cm
\epsffile[220 410 420 540]{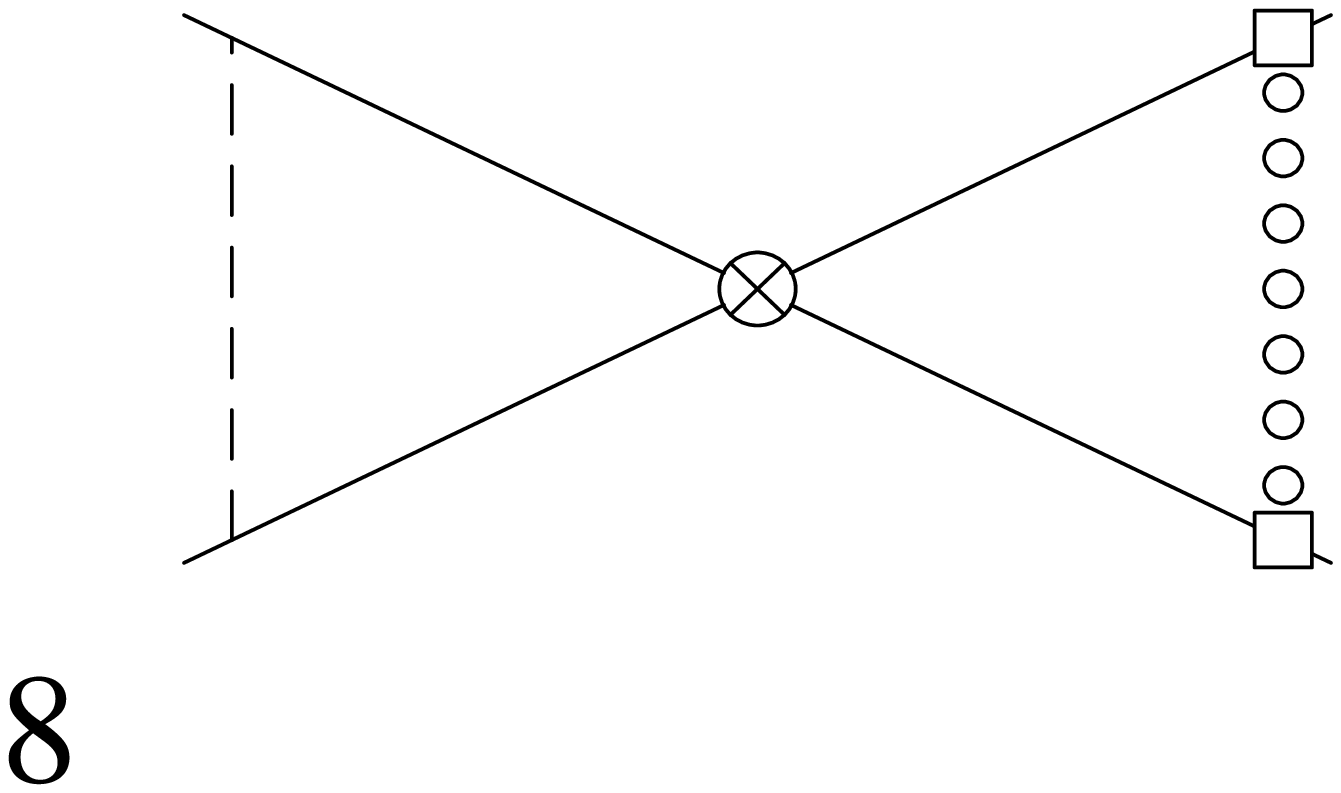}
\hspace{2.2cm}
\leavevmode
\epsfxsize=1.8cm
\epsffile[220 410 420 540]{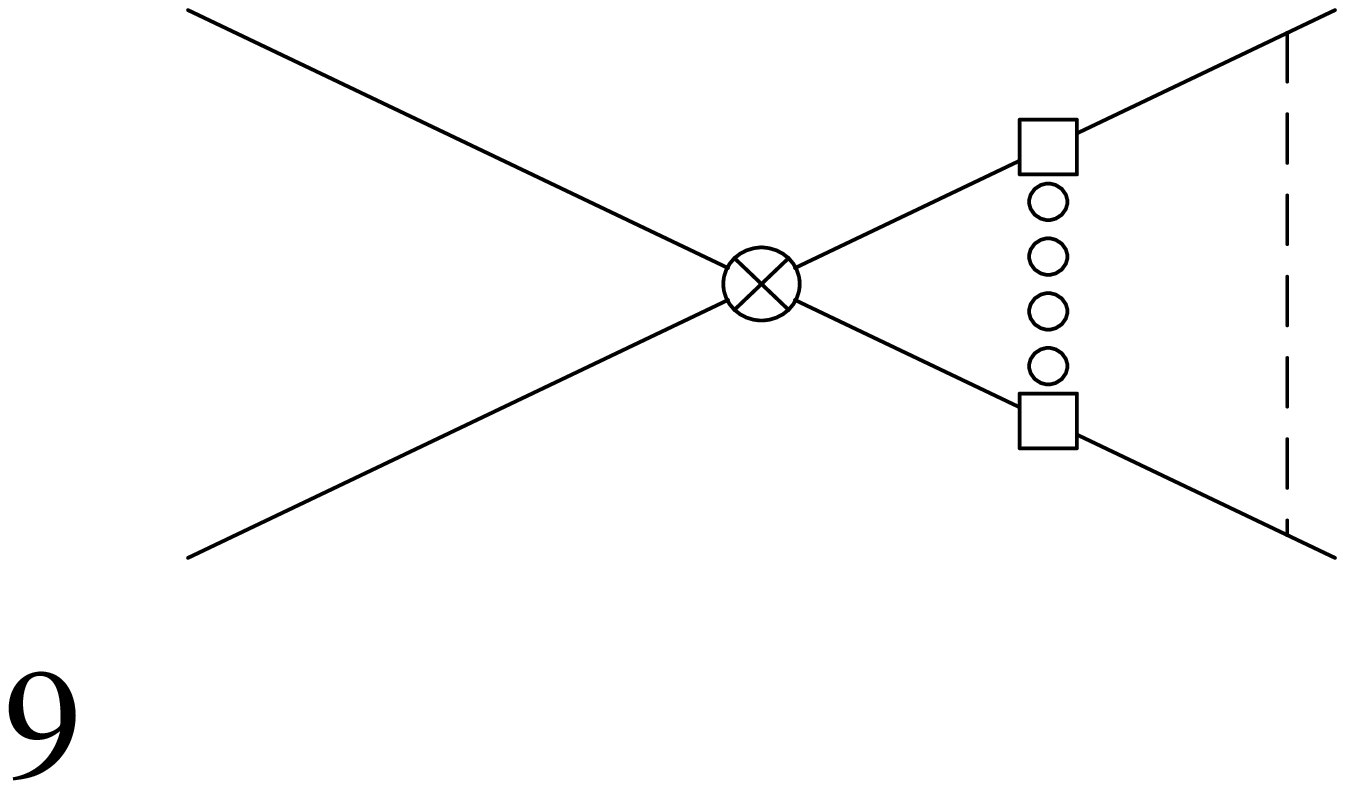}
\\[1.3cm]
\leavevmode
\epsfxsize=1.8cm
\epsffile[220 410 420 540]{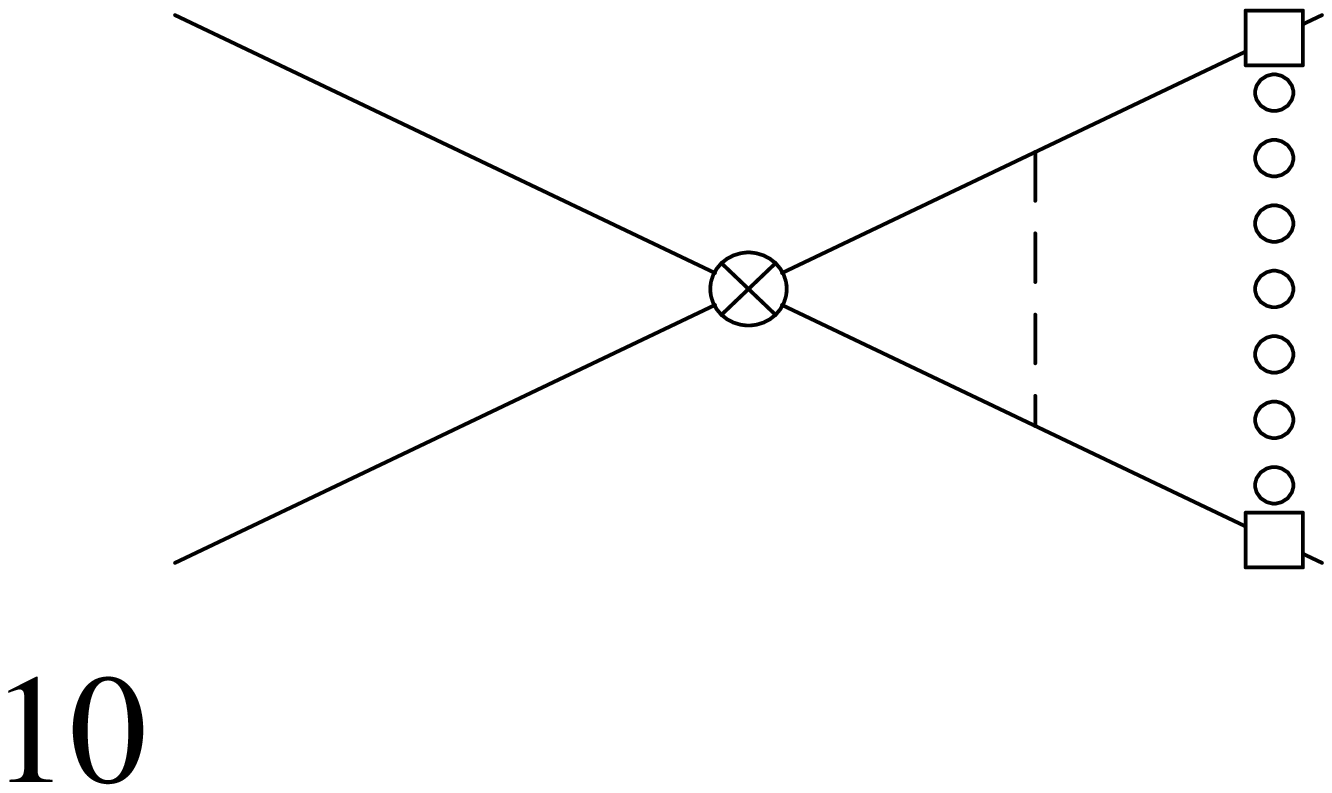}
\hspace{2.2cm}
\leavevmode
\epsfxsize=1.8cm
\epsffile[220 410 420 540]{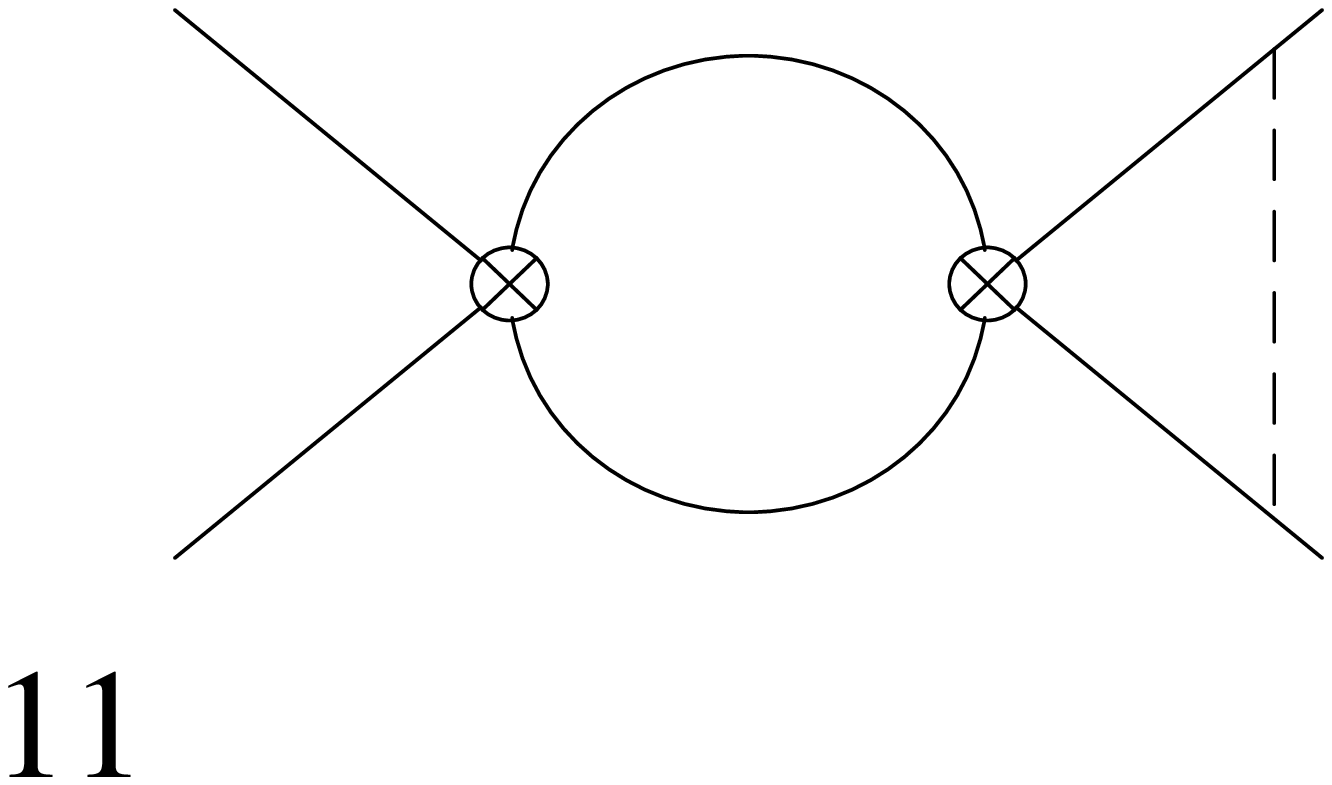}
\hspace{2.2cm}
\leavevmode
\epsfxsize=1.8cm
\epsffile[220 410 420 540]{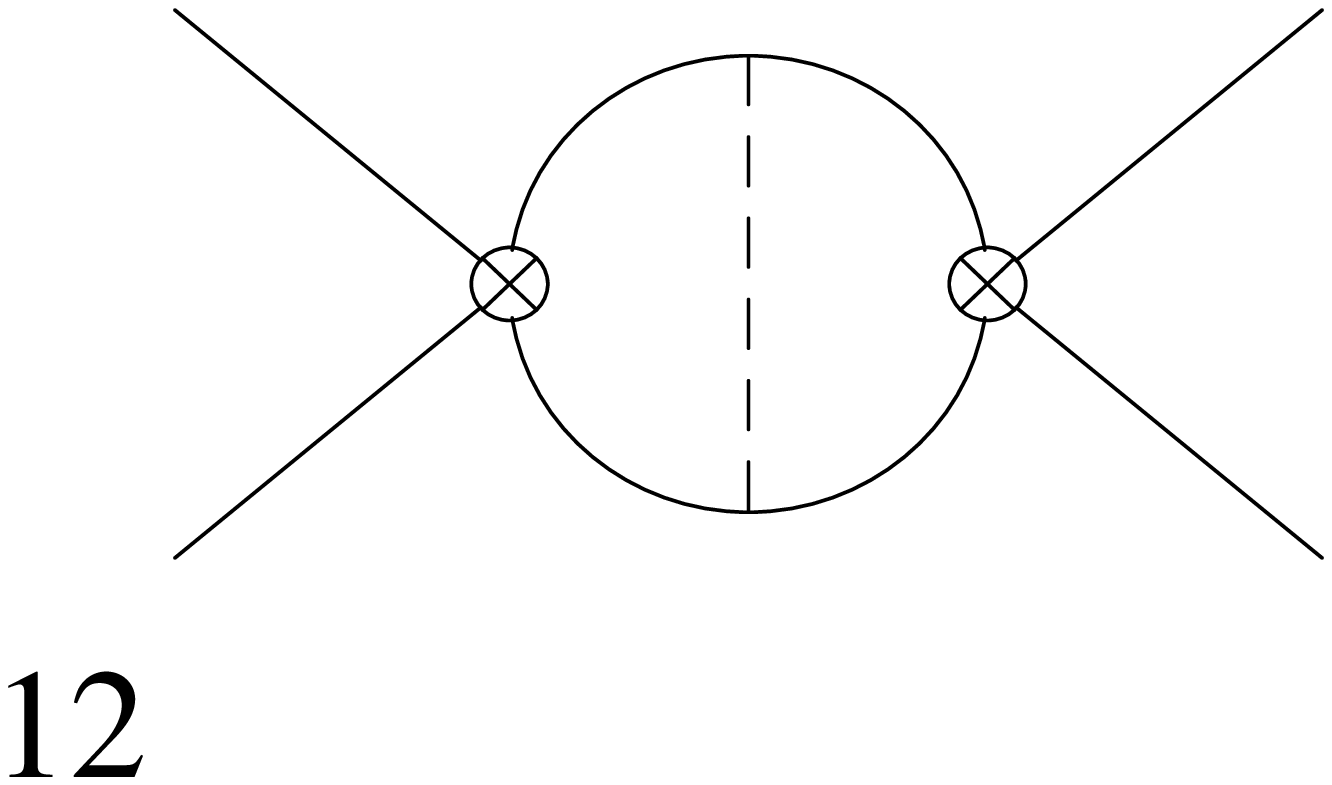}
\vskip  1.5cm
 \caption{\label{figPNRQED2loop} 
Graphical representation of the NRQED single-photon annihilation
scattering diagrams at the two-loop level and at NNLO in the
non-relativistic expansion.
}
 \end{center}
\end{figure}
\begin{figure}[t] 
\begin{center}
\leavevmode
\epsfxsize=2.3cm
\epsffile[220 410 420 540]{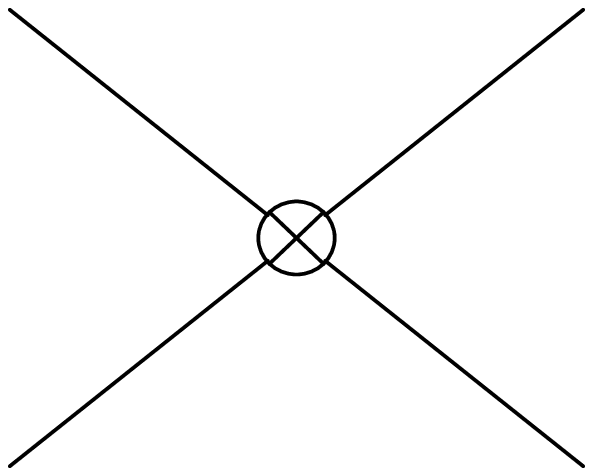}
\hspace{1cm}
\leavevmode
\epsfxsize=2.3cm
\epsffile[220 410 420 540]{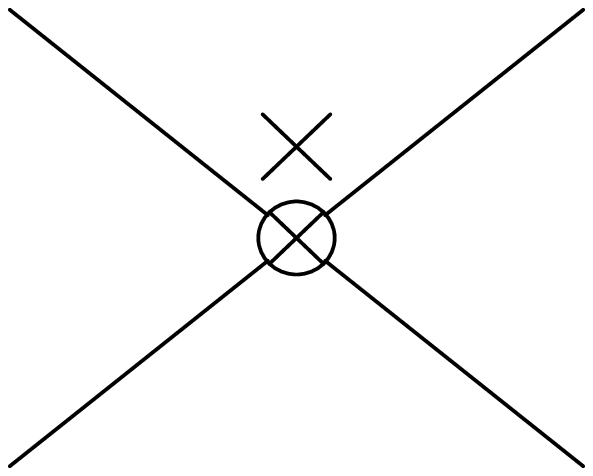}
\hspace{1cm}
\leavevmode
\epsfxsize=2.3cm
\epsffile[220 410 420 540]{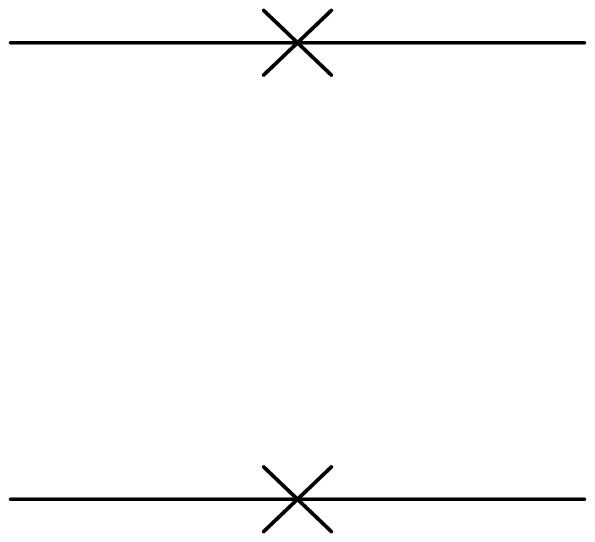}
\\[1.5cm]
\mbox{\hspace{0.2cm}}
$V_{4}$
\mbox{\hspace{2.65cm}}
$V_{4\,\mbox{\tiny der}}$
\mbox{\hspace{2.55cm}}
$\delta H_{\mbox{\tiny kin}}$
\mbox{\hspace{1mm}}\\[.2cm]
\hspace{1cm}
\leavevmode
\epsfxsize=2.3cm
\epsffile[220 410 420 540]{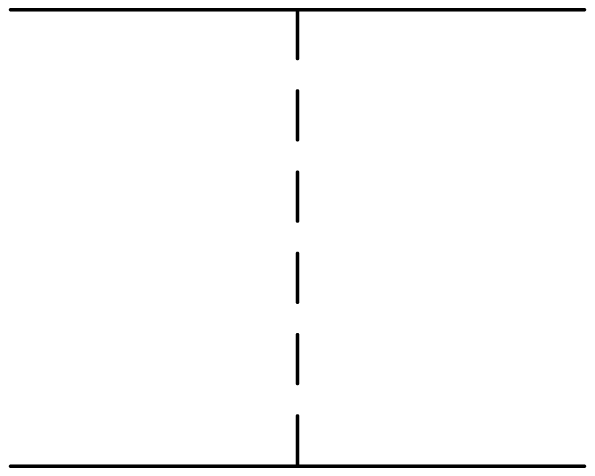}
\hspace{1cm}
\leavevmode
\epsfxsize=2.3cm
\epsffile[220 410 420 540]{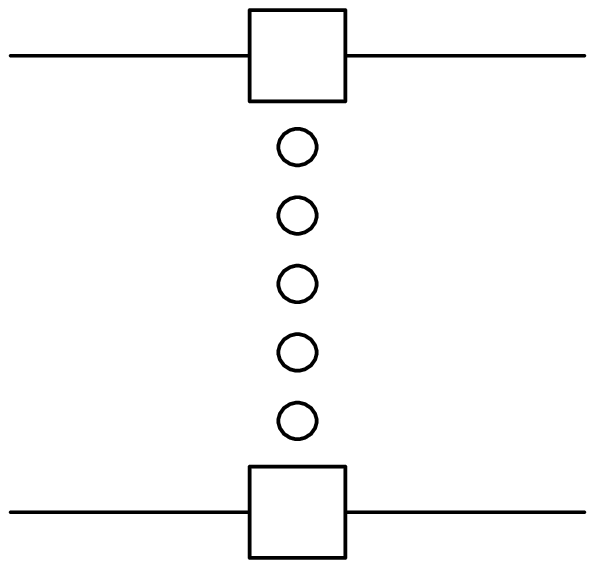}\\[1.5cm]
\mbox{\hspace{1.0cm}}
$V_{\mbox{\tiny Coul}}$
\mbox{\hspace{2.65cm}}
$V_{\mbox{\tiny BF}}$
\mbox{\hspace{.25cm}}
\vskip  .2cm
 \caption{\label{figsymbols}
Symbols describing the NRQED potentials that have to be taken into
account for the matching calculation at NNLO in the non-relativistic
expansion.}
 \end{center}
\end{figure}
To obtain the single-photon scattering amplitude in NRQED we have to
calculate the Feynman diagrams depicted in Figs.~\ref{figPNRQEDborn},
\ref{figPNRQED1loop} and \ref{figPNRQED2loop}, where the various
symbols are defined in Fig.~\ref{figsymbols}. The Feynman diagrams for
the single-photon annihilation contributions of the NRQED scattering
amplitude contain, as the formula for  
$W^{\mbox{\tiny 1-$\gamma$ ann}}$ in Eq.~(\ref{Wfullformula}), at
least one insertion of $V_4$ or $V_{4\mbox{\tiny der}}$. As shown
in the wave equation~(\ref{NNLOSchroedinger}), we can
use time-independent electron--positron propagators also for the NRQED
scattering amplitude. This is possible because all interactions are
instantaneous in time, i.e. the loop integration over the energy
components of the NRQED electron and positron propagators is trivial
by residue (see Eq.~(\ref{residueintegration})).
Because the single-photon annihilation process is only possible for
the electron--positron pair in a ${}^3S_1$ spin triplet state, we only
need to consider the Breit--Fermi potential in the ${}^3S_1$ 
configuration:
\begin{eqnarray}
\tilde V^s_{\mbox{\tiny BF}}(\mbox{\boldmath $p$},
\mbox{\boldmath $q$}) & = &
\frac{5}{3}\,\frac{\pi\,\alpha}{m_e^2}\,
\frac{|{\mbox{\boldmath $p$}}-{\mbox{\boldmath $q$}}|^2}
     {|{\mbox{\boldmath $p$}}-{\mbox{\boldmath $q$}}|^2+\lambda^2}
+\frac{\pi\,\alpha}{m_e^2}\,
\frac{(\mbox{\boldmath $p$}^2-{\mbox{\boldmath $q$}}^2)^2}
   {(|\mbox{\boldmath $p$}-\mbox{\boldmath $q$}|^2+\lambda^2)^2} 
-\frac{\pi\,\alpha}{m_e^2}\,
\frac{|\mbox{\boldmath $p$}+\mbox{\boldmath $q$}|^2}
   {|\mbox{\boldmath $p$}-\mbox{\boldmath $q$}|^2+\lambda^2} 
\,,
\label{BreitFermipotentialsimple}
\end{eqnarray}
where the angular integration it carried out over the angle between
$\mbox{\boldmath $p$}$ and $\mbox{\boldmath $q$}$. We have eliminated
the photon mass in the first term on the RHS of 
Eq.~(\ref{BreitFermipotentialsimple}) because it does not lead
to any infrared divergences. 
An additional
simplification for the NRQED calculation is obtained by replacing the
centre-of-mass energy relative to the $e^+e^-$ threshold,
$E=\sqrt{s}-2m_e$, by the new energy parameter $p_0$, which is defined
as
\begin{equation}
p_0^2 \, \equiv \, \frac{s}{4} - m_e^2
\,. 
\label{p0def}
\end{equation}
The parameter $p_0$ is equal to the relativistic centre-of-mass
three-momentum of the electron/positron in the scattering process. At
NNLO in the non-relativistic expansion we have the relation
\begin{equation}
E \, = \, \frac{p_0^2}{m_e} - \frac{p_0^4}{4\,m_e^3} + \ldots
\,,
\end{equation}
where we keep the term $p_0^2/m_e$ in the LO non-relativistic
electron--positron propagator. In this convention, the NNLO kinetic
energy correction reads
\begin{equation}
\delta H_{\mbox{\tiny kin}}^*(\mbox{\boldmath $p$},\mbox{\boldmath
$q$}) 
\, = \, 
- (2\,\pi)^3\,\delta^{(3)}(\mbox{\boldmath $p$}-\mbox{\boldmath
$q$})\, \frac{\mbox{\boldmath $q$}^4-p_0^4}{4\,m_e^3}
\end{equation} 
and simplifies the form of an insertion of the kinetic energy
correction
\begin{equation}
\int\frac{d^3 \mbox{\boldmath $q$}}{(2\pi)^3}\,
\frac{m_e}{\mbox{\boldmath $p$}^2-p_0^2-i\,\epsilon}\,\Big(-
\delta H_{\mbox{\tiny kin}}^*(\mbox{\boldmath $p$},\mbox{\boldmath
$q$})\,\Big)\, 
\frac{m_e}{\mbox{\boldmath $q$}^2-p_0^2-i\,\epsilon}\,
\, = \,
\frac{m_e}{\mbox{\boldmath $p$}^2-p_0^2-i\,\epsilon}\,
\bigg(
\frac{\mbox{\boldmath $p$}^2+p_0^2}{4\,m_e^3}
\bigg)
\,.
\end{equation}
We emphasize that the introduction of the parameter $p_0$ is just a
technical trick for the matching calculation, which does not affect
the form of $d_1$. Using the ${}^3S_1$ spin average  
\begin{equation}
\frac{1}{3}\,\sum_{J=1,0,-1}
\Big[\,
 (\psi^\dagger{\mbox{\boldmath $\sigma$}}\sigma_2\chi^*)\,
  (\chi^T\sigma_2{\mbox{\boldmath $\sigma$}}\psi)
\,\Big] \, = \, 2
\end{equation}
for the four-fermion operators $V_4$ and $V_{4\mbox{\tiny der}}$ in
Eq.~(\ref{NRQEDLagrangian}), 
we arrive at the following results for the diagrams displayed in
Figs.~\ref{figPNRQEDborn}, \ref{figPNRQED1loop} and
\ref{figPNRQED2loop}  ($D(\mbox{\boldmath $k$})\equiv 
m_e/(\mbox{\boldmath $k$}^2-p_0^2-i\epsilon)$):
\begin{eqnarray}
I_1^{(0)} & = & 
\bigg[\,1+\Big(\frac{\alpha}{\pi}\Big)\,d_1^{(1)} +
\Big(\frac{\alpha}{\pi}\Big)^2\,d_1^{(2)}
\,\bigg]\,
\frac{2\,\pi\,\alpha}{m_e^2}
\,,
\\[2mm]
I_2^{(0)}  & = & 
-\,\frac{8\,\pi\,\alpha}{3\,m_e^2}\,\frac{p_0^2}{m_e^2}
\,,
\\[5mm]
I_1^{(1)}  & = &  2\,i\, 
\bigg[\,1+\Big(\frac{\alpha}{\pi}\Big)\,d_1^{(1)}
\,\bigg]\,
\bigg(\frac{2\,\pi\,\alpha}{m_e^2}\bigg)\,
\int\frac{d^4 k}{(2\pi)^4}\,
\frac{1}{k^0+\frac{p_0^2}{2m_e}-\frac{\mbox{\boldmath $k$}^2}{2m_e}}\,
\frac{1}{k^0-\frac{p_0^2}{2m_e}+\frac{\mbox{\boldmath $k$}^2}{2m_e}}\,
\frac{4\,\pi\,\alpha}
   {(\mbox{\boldmath $k$}-\mbox{\boldmath $p_0$})^2+\lambda^2}
\nonumber
\\[2mm]
& = &  2\, 
\bigg[\,1+\Big(\frac{\alpha}{\pi}\Big)\,d_1^{(1)}
\,\bigg]\,
\bigg(\frac{2\,\pi\,\alpha}{m_e^2}\bigg)\,
\int\frac{d^3 \mbox{\boldmath $k$}}{(2\pi)^3}\,
D(\mbox{\boldmath $k$})\,
\frac{4\,\pi\,\alpha}
   {(\mbox{\boldmath $k$}-\mbox{\boldmath $p_0$})^2+\lambda^2}
\nonumber
\\[2mm]
& = &   
\bigg[\,1+\Big(\frac{\alpha}{\pi}\Big)\,d_1^{(1)}
\,\bigg]\,
\frac{2\,\alpha^2}{m_e\,p_0}\,
\bigg[\,
\frac{\pi^2}{2}+i\,\pi\,\ln\Big(\frac{2\,p_0}{\lambda}\Big)
\,\bigg]
\,,
\label{residueintegration}
\\[2mm]
I_2^{(1)}  & = &  -2\,
\bigg(\frac{4\,\pi\,\alpha}{3\,m_e^2}\bigg)\,
\int\frac{d^3 \mbox{\boldmath $k$}}{(2\pi)^3}\,
\bigg(\frac{\mbox{\boldmath $k$}^2+p_0^2}{m_e^2}\bigg)\,
D(\mbox{\boldmath $k$})\,
\frac{4\,\pi\,\alpha}
   {(\mbox{\boldmath $k$}-\mbox{\boldmath $p_0$})^2+\lambda^2}
\nonumber
\\[2mm]
& = & -\,\frac{4\,\alpha^2}{3\,m_e^2}\,
\bigg[\,
\frac{4\,\Lambda}{m_e} + 
\frac{p_0\,\pi^2}{m_e} +
2\,i\,\pi\,\frac{p_0}{m_e}\,\ln\Big(\frac{2\,p_0}{\lambda}\Big)
\,\bigg]
\,,
\\[2mm]
I_3^{(1)}  & = &   2\,
\bigg(\frac{2\,\pi\,\alpha}{m_e^2}\bigg)\,
\int\frac{d^3 \mbox{\boldmath $k$}}{(2\pi)^3}\,
D(\mbox{\boldmath $k$})\,
\bigg(\frac{\mbox{\boldmath $k$}^2+p_0^2}{4\,m_e^2}\bigg)\,
\frac{4\,\pi\,\alpha}
   {(\mbox{\boldmath $k$}-\mbox{\boldmath $p_0$})^2+\lambda^2}
\nonumber
\\[2mm]
& = & \frac{\alpha^2}{2\,m_e^2}\,
\bigg[\,
\frac{4\,\Lambda}{m_e} + 
\frac{p_0\,\pi^2}{m_e} +
2\,i\,\pi\,\frac{p_0}{m_e}\,\ln\Big(\frac{2\,p_0}{\lambda}\Big)
\,\bigg]
\,,
\\[2mm]
I_4^{(1)}  & = &  2\,
\bigg(\frac{2\,\pi\,\alpha}{m_e^2}\bigg)\,
\int\frac{d^3 \mbox{\boldmath $k$}}{(2\pi)^3}\,
D(\mbox{\boldmath $k$})\,
\Big(
-\tilde V^s_{\mbox{\tiny BF}}(\mbox{\boldmath $k$},
\mbox{\boldmath $p_0$})\,
\Big)
\nonumber
\\[2mm]
& = & \frac{\alpha^2}{m_e^2}\,
\bigg[\,
-\frac{10\,\Lambda}{3\,m_e} + 
\frac{p_0\,\pi^2}{m_e} +
2\,i\,\pi\,\frac{p_0}{m_e}\,
  \bigg(\,-\frac{4}{3}+\ln\Big(\frac{2\,p_0}{\lambda}\Big)
  \,\bigg)
\,\bigg]
\,,
\\[2mm]
I_5^{(1)}  & = &  
-\,\bigg(\frac{2\,\pi\,\alpha}{m_e^2}\bigg)^2\,
\int\frac{d^3 \mbox{\boldmath $k$}}{(2\pi)^3}\,
D(\mbox{\boldmath $k$})
\, = \, 
-\frac{\alpha^2}{m_e^2}\,
\bigg[\,
\frac{2\,\Lambda}{m_e} + 
i\,\pi\,\frac{p_0}{m_e}
\,\bigg]
\,,
\\[5mm]
I_1^{(2)}  & = & 
\bigg(\frac{m_e^2}{8\,\pi\,\alpha}\bigg)\,
\Big[\, I_1^{(1)} \,\Big]^2
\,,
\\[2mm]
I_2^{(2)}  & = &  2\,
\bigg(\frac{2\,\pi\,\alpha}{m_e^2}\bigg)\,
\int\frac{d^3 \mbox{\boldmath $k_1$}}{(2\pi)^3}\,
\int\frac{d^3 \mbox{\boldmath $k_2$}}{(2\pi)^3}\,
D(\mbox{\boldmath $k_1$})\,
\frac{4\,\pi\,\alpha}
   {(\mbox{\boldmath $k_1$}-\mbox{\boldmath $k_2$})^2+\lambda^2}\,
D(\mbox{\boldmath $k_2$})\,
\frac{4\,\pi\,\alpha}
   {(\mbox{\boldmath $k_2$}-\mbox{\boldmath $p_0$})^2+\lambda^2}
\nonumber
\\[2mm]
& = & 
\frac{\pi\,\alpha^3}{2\,p_0^2}\,
\bigg[\,
\frac{\pi^2}{12}-\ln^2\Big(\frac{2\,p_0}{\lambda}\Big)
+i\,\pi\,\ln\Big(\frac{2\,p_0}{\lambda}\Big)
\,\bigg]
\,,
\\[2mm]
I_3^{(2)}  & = &  
-\bigg(\frac{4\,\pi\,\alpha}{3\,m_e^2}\bigg)\,
\int\frac{d^3 \mbox{\boldmath $k_1$}}{(2\pi)^3}\,
\int\frac{d^3 \mbox{\boldmath $k_2$}}{(2\pi)^3}\,
\frac{4\,\pi\,\alpha}
   {(\mbox{\boldmath $p_0$}-\mbox{\boldmath $k_1$})^2+\lambda^2}\,
D(\mbox{\boldmath $k_1$})\,
\bigg(\,\frac{\mbox{\boldmath $k_1$}^2+\mbox{\boldmath $k_2$}^2}
             {m_e^2}\,\bigg)\,
D(\mbox{\boldmath $k_2$})\,
\frac{4\,\pi\,\alpha}
   {(\mbox{\boldmath $k_2$}-\mbox{\boldmath $p_0$})^2+\lambda^2}
\nonumber
\\[2mm]
& = & 
-\frac{2\,\alpha^3}{3\,\pi\,m_e\,p_0}\,
\bigg[\,
\frac{\pi^2}{2}
+i\,\pi\,\ln\Big(\frac{2\,p_0}{\lambda}\Big)
\,\bigg]\,
\bigg[\,
\frac{4\,\Lambda}{m_e} + 
\frac{p_0\,\pi^2}{2\,m_e} +
i\,\pi\,\frac{p_0}{m_e}\,\ln\Big(\frac{2\,p_0}{\lambda}\Big)
\,\bigg]
\,,
\\[2mm]
I_4^{(2)}  & = &  
-2\,\bigg(\frac{4\,\pi\,\alpha}{3\,m_e^2}\bigg)\,
\int\frac{d^3 \mbox{\boldmath $k_1$}}{(2\pi)^3}\,
\int\frac{d^3 \mbox{\boldmath $k_2$}}{(2\pi)^3}\,
D(\mbox{\boldmath $k_1$})\,
\bigg(\,\frac{\mbox{\boldmath $p_0$}^2+\mbox{\boldmath $k_1$}^2}
             {m_e^2}\,\bigg)\,
\frac{4\,\pi\,\alpha}
   {(\mbox{\boldmath $k_1$}-\mbox{\boldmath $k_2$})^2+\lambda^2}\,
D(\mbox{\boldmath $k_2$})\,
\frac{4\,\pi\,\alpha}
   {(\mbox{\boldmath $k_2$}-\mbox{\boldmath $p_0$})^2+\lambda^2}
\nonumber
\\[2mm]
& = & 
-\frac{2\,\alpha^3}{3\,m_e\,p_0}\,
\bigg[\,
\frac{2\,\Lambda}{m_e}\,\bigg( \pi
+2\,i\,\ln\Big(\frac{2\,p_0}{\lambda}\Big)
\bigg)
-\frac{2\,p_0\,\pi}{m_e}\,\bigg(
1-\frac{\pi^2}{24}+
\frac{1}{2}\ln^2\Big(\frac{2\,p_0}{\lambda}\Big)
-i\,\frac{\pi}{2}\,\ln\Big(\frac{2\,p_0}{\lambda}\Big) 
\bigg)
\,\bigg]
\,,
\\[2mm]
I_5^{(2)}  & = &  
\bigg(\frac{m_e^2}{4\,\pi\,\alpha}\bigg)\,
\Big[\, I_1^{(1)} \,\Big]\,\Big[\, I_3^{(1)} \,\Big]
\,,
\\[2mm]
I_6^{(2)}  & = &   2\,
\bigg(\frac{2\,\pi\,\alpha}{m_e^2}\bigg)\,
\int\frac{d^3 \mbox{\boldmath $k_1$}}{(2\pi)^3}\,
\int\frac{d^3 \mbox{\boldmath $k_2$}}{(2\pi)^3}\,
D(\mbox{\boldmath $k_1$})\,
\bigg(\frac{\mbox{\boldmath $k_1$}^2+p_0^2}{4\,m_e^2}\bigg)\,
\frac{4\,\pi\,\alpha}
   {(\mbox{\boldmath $k_1$}-\mbox{\boldmath $k_2$})^2+\lambda^2}\,
D(\mbox{\boldmath $k_2$})\,
\frac{4\,\pi\,\alpha}
   {(\mbox{\boldmath $k_2$}-\mbox{\boldmath $p_0$})^2+\lambda^2}
\nonumber
\\[2mm]
& = & 
\frac{\alpha^3}{4\,m_e\,p_0}\,
\bigg[\,
\frac{2\,\Lambda}{m_e}\,\bigg( \pi
+2\,i\,\ln\Big(\frac{2\,p_0}{\lambda}\Big)
\bigg)
-\frac{2\,p_0\,\pi}{m_e}\,\bigg(
1-\frac{\pi^2}{24}+
\frac{1}{2}\ln^2\Big(\frac{2\,p_0}{\lambda}\Big)
-i\,\frac{\pi}{2}\,\ln\Big(\frac{2\,p_0}{\lambda}\Big) 
\bigg)
\,\bigg]
\,,
\\[2mm]
I_7^{(2)}  & = &   2\,
\bigg(\frac{2\,\pi\,\alpha}{m_e^2}\bigg)\,
\int\frac{d^3 \mbox{\boldmath $k_1$}}{(2\pi)^3}\,
\int\frac{d^3 \mbox{\boldmath $k_2$}}{(2\pi)^3}\,
D(\mbox{\boldmath $k_1$})\,
\frac{4\,\pi\,\alpha}
   {(\mbox{\boldmath $k_1$}-\mbox{\boldmath $k_2$})^2+\lambda^2}\,
D(\mbox{\boldmath $k_2$})\,
\bigg(\frac{\mbox{\boldmath $k_2$}^2+p_0^2}{4\,m_e^2}\bigg)\,
\frac{4\,\pi\,\alpha}
   {(\mbox{\boldmath $k_2$}-\mbox{\boldmath $p_0$})^2+\lambda^2}
\nonumber
\\[2mm]
& = & 
\frac{\pi\,\alpha^3}{m_e^2}\,
\bigg[\,
\frac{\pi^2}{48}
+1 - \ln\Big(\frac{2\,p_0}{\Lambda}\Big)
- \frac{1}{4}\,\ln^2\Big(\frac{2\,p_0}{\lambda}\Big)
+i\,\frac{\pi}{2}\,\bigg(
1+\frac{1}{2}\,\ln\Big(\frac{2\,p_0}{\lambda}\Big)
\bigg)
\,\bigg]
\,,
\\[2mm]
I_8^{(2)}  & = &   
\bigg(\frac{m_e^2}{4\,\pi\,\alpha}\bigg)\,
\Big[\, I_1^{(1)} \,\Big]\,\Big[\, I_4^{(1)} \,\Big]
\,,
\\[2mm]
I_9^{(2)}  & = & 2\,
\bigg(\frac{2\,\pi\,\alpha}{m_e^2}\bigg)\,
\int\frac{d^3 \mbox{\boldmath $k_1$}}{(2\pi)^3}\,
\int\frac{d^3 \mbox{\boldmath $k_2$}}{(2\pi)^3}\,
D(\mbox{\boldmath $k_1$})\, 
\Big(
-\tilde V^s_{\mbox{\tiny BF}}(\mbox{\boldmath $k_1$},
\mbox{\boldmath $k_2$})\,
\Big)\,
D(\mbox{\boldmath $k_2$})\,
\frac{4\,\pi\,\alpha}
   {(\mbox{\boldmath $k_2$}-\mbox{\boldmath $p_0$})^2+\lambda^2}
\nonumber
\\[2mm]
& = & 
\frac{\alpha^3}{m_e\,p_0}\,
\bigg[\,
\frac{\pi\,p_0}{m_e}\,\bigg(
\frac{\pi^2}{24}+1-\frac{1}{2}\ln 2
-2\, \ln\Big(\frac{2\,p_0}{\Lambda}\Big)
+\frac{11}{6}\, \ln\Big(\frac{2\,p_0}{\lambda}\Big)
-\frac{1}{2}\, \ln^2\Big(\frac{2\,p_0}{\lambda}\Big)
\bigg)
\nonumber
\\[2mm] & &
\hspace{2cm}
-\frac{5\,\Lambda}{6\,m_e}\,\bigg(
\pi+2\,i\,\ln\Big(\frac{2\,p_0}{\lambda}\Big)
\bigg)
+i\,\frac{\pi^2\,p_0}{2\,m_e}\,\bigg(
\frac{1}{6}+ \ln\Big(\frac{2\,p_0}{\lambda}\Big)
\bigg)
\,\bigg]
\,,
\\[2mm]
I_{10}^{(2)}  & = & 2\,
\bigg(\frac{2\,\pi\,\alpha}{m_e^2}\bigg)\,
\int\frac{d^3 \mbox{\boldmath $k_1$}}{(2\pi)^3}\,
\int\frac{d^3 \mbox{\boldmath $k_2$}}{(2\pi)^3}\,
D(\mbox{\boldmath $k_1$})\,
\frac{4\,\pi\,\alpha}
   {(\mbox{\boldmath $k_1$}-\mbox{\boldmath $k_2$})^2+\lambda^2}\,
D(\mbox{\boldmath $k_2$})\,
\nonumber
\\[2mm] & &
\hspace{5cm}
\times\,
\bigg(\,
2\,\frac{\pi\,\alpha}{m_e^2}\,
\frac{\mbox{\boldmath $k_2$}^2+p_0^2}
   {(\mbox{\boldmath $k_2$}-\mbox{\boldmath $p_0$})^2+\lambda^2} -
\frac{11}{3}\,\frac{\pi\,\alpha}{m_e^2}\,
\bigg)
\nonumber
\\[2mm]
& = & 
\frac{\pi\,\alpha^3}{2\,m_e^2}\,\bigg[\,
\frac{\pi^2}{12}+4+\ln 2
+\frac{10}{3}\,\ln\Big(\frac{2\,p_0}{\Lambda}\Big)
-\ln\Big(\frac{2\,p_0}{\lambda}\Big)
-\ln^2\Big(\frac{2\,p_0}{\lambda}\Big)
+ i\,\pi\bigg(
-\frac{7}{6}+\ln\Big(\frac{2\,p_0}{\lambda}\Big)
\bigg)
\,\bigg]
\,,
\\[2mm]
I_{11}^{(2)}  & = & 
\bigg(\frac{m_e^2}{2\,\pi\,\alpha}\bigg)\,
\Big[\, I_1^{(1)} \,\Big]\,\Big[\, I_5^{(1)} \,\Big]
\,,
\\[2mm]
I_{12}^{(2)}  & = &  
-\,\bigg(\frac{2\,\pi\,\alpha}{m_e^2}\bigg)^2\,
\int\frac{d^3 \mbox{\boldmath $k_1$}}{(2\pi)^3}\,
\int\frac{d^3 \mbox{\boldmath $k_2$}}{(2\pi)^3}\,
D(\mbox{\boldmath $k_1$})\,
\frac{4\,\pi\,\alpha}
   {(\mbox{\boldmath $k_1$}-\mbox{\boldmath $k_2$})^2+\lambda^2}\,
D(\mbox{\boldmath $k_2$})\,
\, = \, 
\frac{\pi\,\alpha^3}{m_e^2}\,
\bigg[\,
\ln\Big(\frac{2\,p_0}{\Lambda}\Big)
-i\,\frac{\pi}{2}
\,\bigg]
\,.
\end{eqnarray}
The upper index of the functions $I_j^{(i)}$ corresponds to the
power of the fine structure constant of the diagrams and
the lower index to the numeration given in Figs.~\ref{figPNRQEDborn},
\ref{figPNRQED1loop} and \ref{figPNRQED2loop}.
Combinatorial factors are taken into
account. We note that all the above results have been given in the
limit  $\lambda\ll p_0\ll m_e$ and that only the powers of $p_0$
relevant at NNLO have been kept.
A collection of integrals that have been useful in determining the
results given above is presented in Appendix.~\ref{appendixshortdistance}.
The full NRQED amplitude reads
\begin{eqnarray}
{\cal{A}}_{\mbox{\tiny NRQED}}^{\mbox{\tiny 1-$\gamma$ ann}} & = &
\sum\limits_{i=1}^{2}\,I^{(0)}_i +
\sum\limits_{i=1}^{5}\,I^{(1)}_i +
\sum\limits_{i=1}^{12}\,I^{(2)}_i
\,.
\label{fullPNRQEDamplitude}
\end{eqnarray}

The elastic single-photon annihilation amplitude in full QED can be
obtained from the electromagnetic form factors, which parametrize the
radiative corrections to the electromagnetic vertex, and the photon
vacuum polarization function. The electromagnetic form factors $F_1$
(Dirac) and $F_2$ (Pauli) are defined through
\begin{equation}
\bar{u}(p^\prime)\,\Lambda_\mu^{em}\,v(p) \,=\,
i\,e\,\bar{u}(p^\prime)\,\bigg[\,
\gamma_\mu\,F_1(q^2) 
+ \frac{i}{2\,M}\,\sigma_{\mu\nu}\,q^\nu\,F_2(q^2)
\,\bigg]\,v(p)
\,,
\end{equation}
for the $e^+e^-$ production vertex, where $q=p+p^\prime$ and
$\sigma_{\mu\nu}=\frac{i}{2}\,[\,\gamma_\mu,\gamma_\nu\,]$.
We need the form factors in the limit $\lambda\ll p_0\ll m_e$.
The one-loop contributions have been known for a long time for all
momenta~\cite{Kallensabry1,Schwinger1}, whereas the two-loop
contributions have been 
calculated in the desired limit in Ref.~\cite{Hoang2}. 
Parametrizing
the loop corrections to the form factors as
\begin{eqnarray}
F_1(q^2) & = & 1 \, + \, \Big(\frac{\alpha}{\pi}\Big)\,F_1^{(1)}(q^2) 
\, + \,
\Big(\frac{\alpha}{\pi}\Big)^2\,F_1^{(2)}(q^2) \, + \, \cdots
\,,\nonumber\\[2mm]
F_2(q^2) & = & \hspace{0.8cm} \Big(\frac{\alpha}{\pi}\Big)\,
F_2^{(1)}(q^2) \, + \,
\Big(\frac{\alpha}{\pi}\Big)^2\,F_2^{(2)}(q^2) \, + \, \cdots
\,,
\end{eqnarray}
and using the energy parameter $p_0$ (Eq.~(\ref{p0def})) the results
for the form factors in the threshold region at NNLO in the
non-relativistic expansion read ($\lambda\ll p_0\ll m_e$):
\begin{eqnarray}
F_1^{(1)}(q^2) & = &
i\,\frac{\pi\,m_e}{2\,p_0}\,\bigg[\,\ell-\frac{1}{2} \,\bigg] 
- \frac{3}{2} 
+  i\,\frac{3\,\pi\,p_0}{4\,m_e}\,\bigg[\,\ell-\frac{5}{6} \,\bigg] 
+ {\cal{O}}\Big(\frac{p_0^2}{m_e^2}\Big)
\,,
\label{F1oneloop}
\\[2mm]
F_1^{(2)}(q^2) & = &
\frac{\pi^2\,m_e^2}{8\,p_0^2}\,\bigg[\,
-\ell^2+\ell-\frac{\pi^2}{6}-\frac{1}{3}
\,\bigg]
- i\,\frac{\pi\,m_e}{4\,p_0}\,\bigg[\,
3\,\ell -1 
\,\bigg] 
\nonumber
\\[2mm] & &
- \bigg[\,
\frac{\pi^4}{16}
+\frac{3\,\pi^2}{8}\,\bigg(\,
\ell^2 - \frac{4}{3}\,\ell+ 
\frac{46}{45}\,\ln\Big(-i\,\frac{p_0}{m_e}\Big)
+\frac{7}{15}\,\ln 2 
+\frac{2729}{1350}
\,\bigg)
\nonumber
\\[2mm] & &
\hspace{1cm}
+\frac{9}{80}\,\bigg(9\,\zeta_3-43\bigg)
\,\bigg] 
+ {\cal{O}}\Big(\frac{p_0}{m_e}\Big)
\,,
\label{F1twoloop}
\\[2mm]
F_2^{(1)}(q^2) & = &
i\,\frac{\pi\,m_e}{4\,p_0} - \frac{1}{2} - 
i\,\frac{\pi\,p_0}{8\,m_e} \,
+ {\cal{O}}\Big(\frac{p_0^2}{m_e^2}\Big)
\,,
\label{F2oneloop}
\\[2mm]
F_2^{(2)}(q^2) & = &
\frac{\pi^2\,m_e^2}{8\,p_0^2}\,\bigg[\,
-\ell+\frac{1}{3}
\,\bigg]
- i\,\frac{\pi\,m_e}{4\,p_0}\,\bigg[\,
\ell + 1 
\,\bigg] 
\nonumber
\\[2mm] & &
+ \bigg[\,
\frac{\pi^2}{8}\,\bigg(\,
- \ell+ 
\frac{2}{5}\,\ln\Big(-i\,\frac{p_0}{m_e}\Big)
+\frac{101}{15}\,\ln 2 
-\frac{1621}{450}
\,\bigg)
\nonumber
\\[2mm] & &
\hspace{1cm}
+\frac{1}{80}\,\bigg(41\,\zeta_3 + \frac{347}{9}\bigg)
\,\bigg] 
+ {\cal{O}}\Big(\frac{p_0}{m_e}\Big)
\,,
\label{F2twoloop}
\\[2mm]
\end{eqnarray}
where
\begin{equation}
\ell \, \equiv \, \ln\Big(-\frac{2\,i\,p_0}{\lambda}\Big)
\,.
\end{equation}
The photon vacuum polarization function $\Pi$ is defined as
\begin{equation}
(q^2\,g^{\mu\nu}-q^\mu\,q^\nu)\,\Pi(q^2)\, \equiv \,
i\,\int d^4x\, e^{i q x}\,\langle\,0\,|\,
T\,j^\mu(x)\,j^\nu(0)\,
|\,0\,\rangle
\,,
\end{equation}
where $j^\nu$ is the electromagnetic current.
The one- and two-loop contributions to $\Pi$ are also known from
Refs.~\cite{Kallensabry1,Schwinger1} for all values of $q^2$ and read,
expanded up to NNLO in the non-relativistic expansion,
\begin{eqnarray}
\Pi(q^2) & = & \Big(\frac{\alpha}{\pi}\Big)\,\Pi^{(1)}(q^2) 
\, + \,
\Big(\frac{\alpha}{\pi}\Big)^2\,\Pi^{(2)}(q^2) \, + \, \cdots
\,,
\end{eqnarray}
with
\begin{eqnarray}
\Pi^{(1)}(q^2) & \stackrel{p_0\ll m_e}{=} &
\frac{8}{9} + i\,\frac{\pi\,p_0}{2\,m_e}
+ {\cal{O}}\Big(\frac{p_0^2}{m_e^2}\Big)
\,,
\label{Poneloop}
\\[2mm]
\Pi^{(2)}(q^2) & \stackrel{p_0\ll m_e}{=} &
-\frac{\pi^2}{2}\,\bigg(
\ln\Big(-i\,\frac{p_0}{m_e}\Big) 
-\frac{11}{16}+\frac{3}{2}\,\ln 2
\bigg)
-\frac{21}{8}\,\zeta_3 
+\frac{3}{4}
+ {\cal{O}}\Big(\frac{p_0}{m_e}\Big)
\,.
\label{Ptwoloop}
\end{eqnarray}
Including all effects up to NNLO in $p_0/m_e$ and taking the ${}^3S_1$
spin average, the QED amplitude reads
\begin{eqnarray}
{\cal{A}}_{\mbox{\tiny QED}}^{\mbox{\tiny 1-$\gamma$ ann}} & = &
\bigg(\frac{2\,\alpha\,\pi}{m_e^2}\bigg)\,
\bigg(1-\frac{p_0^2}{m_e^2}\bigg)\,
\frac{1}{1+\Pi(q^2)}\,
\bigg[\,
\bigg(1-\frac{p_0^2}{6\,m_e^2}\bigg)\,F_1(q^2) +
\bigg(1+\frac{p_0^2}{6\,m_e^2}\bigg)\,F_2(q^2)
\,\bigg]^2
\,.
\label{fullQEDamplitude}
\end{eqnarray}
The short-distance coefficient $d_1$ is determined by 
requiring equality of all the terms up to order $\alpha^3$ and NNLO in
$p_0/m_e$ in the QED and the NRQED amplitudes in
Eqs.~(\ref{fullQEDamplitude}) and (\ref{fullPNRQEDamplitude}).
The result for $d_1$ reads
\begin{eqnarray}
d_1 & = & 1 + \Big(\frac{\alpha}{\pi}\Big)\,d_1^{(1)} 
\, + \,
\Big(\frac{\alpha}{\pi}\Big)^2\,d_1^{(2)} \, + \, \cdots
\,,
\label{d1final}
\end{eqnarray}
where
\begin{eqnarray}
d_1^{(1)} & = &
\frac{13}{3}\,\frac{\Lambda}{m_e}-\frac{44}{9}
\,,
\label{d1oneloop}
\\[2mm]
d_1^{(2)} & = &
- \frac{\pi^2}{6}\,\ln\Big(\frac{\Lambda}{m_e}\Big) 
+ \frac{13}{8}\,\zeta_3 
- \frac{1483\,\pi^2}{288} 
+ \frac{9\,\pi^2}{4}\,\ln 2 
+ \frac{1477}{81} 
\,.
\label{d1twoloop}
\end{eqnarray}

\vspace{1.5cm}
\section{The Bound State Calculation}
\label{sectionboundstate}
In the final step we have to evaluate the RHS of
Eq.~(\ref{Wfullformula}). It is convenient to perform this calculation
also in momentum space representation. In this
representation the normalized $n=1$, ${}^3S_1$ positronium bound state
wave function in the non-relativistic limit reads
\begin{equation} 
\phi_0({\mbox{\boldmath $p$}}) 
\, \equiv \,
\langle \mbox{\boldmath $p$}\, |\, 1 {}^3S_1 \,\rangle
\, = \,
\frac{8\,\sqrt{\pi}\,\gamma^{5/2}}
{({\mbox{\boldmath $p$}}^2+\gamma^2)^2} 
\,,
\end{equation}
where 
\begin{equation}
\gamma \, \equiv \, \frac{m_e\,\alpha}{2}
\,.
\end{equation}
We also need a momentum space expression for the sum over intermediate Coulomb
states in the third and fourth terms on the RHS of
Eq.~(\ref{Wfullformula}). This sum is just the Green function of the
non-relativistic Coulomb problem, where the $n=1$, ${}^3S_1$ ground
state pole is subtracted and $E=E_0=-m_e\alpha^2/4$. A compact momentum
space integral representation for the Coulomb Green function has been
determined by Schwinger~\cite{Schwinger2}:
\begin{eqnarray}
\lefteqn{
\langle \mbox{\boldmath $p$}\, | \,
  \sum\hspace{-5.5mm}\int\limits_{l}\,
  \frac{|\,l\,\rangle\,\langle \,l\,|}{E_l-E-i\,\epsilon}
\, | \, \mbox{\boldmath $q$}\,\rangle
\, = \,
\frac{(2\,\pi)^3\,
  \delta^{(3)}(\mbox{\boldmath $p$}-\mbox{\boldmath $q$})\,m_e}
{\mbox{\boldmath $p$}^2-m_e\,E-i\,\epsilon} 
+ \frac{m_e}{\mbox{\boldmath $p$}^2-m\,E-i\,\epsilon}\,
  \frac{4\,\pi\,\alpha}
     {(\mbox{\boldmath $p$}-\mbox{\boldmath $q$})^2}\,
  \frac{m_e}{\mbox{\boldmath $q$}^2-m\,E-i\,\epsilon}
} 
\label{SchwingerGreenfunction}
\\[2mm] & &
- \frac{4\,\pi\,\alpha\,m_e}{\mbox{\boldmath
  $p$}^2-m\,E-i\,\epsilon}\,
\int\limits_0^1 dx\,\frac{i\,\eta\,x^{-i\,\eta}}
  {
  (\mbox{\boldmath $p$}-\mbox{\boldmath $q$})^2\,x - 
  \frac{1}{4\,m_e\,E}\,(\mbox{\boldmath $p$}^2-m_e\,E-i\,\epsilon)\,
  (\mbox{\boldmath $q$}^2-m_e\,E-i\,\epsilon)\,(1-x)^2
  }\,
\frac{m_e}{\mbox{\boldmath $q$}^2}
\,,
\nonumber
\end{eqnarray}
where
\begin{eqnarray}
i\,\eta \, = \, \frac{\alpha}{2}\,\sqrt{\frac{m_e}{-E-i\,\epsilon}}
\,.
\end{eqnarray} 
Taking the limit $E\to E_0$ the third term on the RHS of
Eq.~(\ref{SchwingerGreenfunction}) develops the $n=1$, ${}^3S_1$ pole,
$|\phi_0({\mbox{\boldmath $p$}})|^2/(E_0-E)$. After subtraction of
this pole, one finds that the  momentum space representation of the sum over
intermediate states in Eq.~(\ref{Wfullformula}) can be written as:
\begin{eqnarray}
\lefteqn{
\langle \mbox{\boldmath $p$}\, | \,
  \sum\hspace{-7.5mm}\int\limits_{l\ne 1{}^3S_1}\,
  \frac{|\,l\,\rangle\,\langle \,l\,|}{E_0-E_l+i\,\epsilon}
\, | \, \mbox{\boldmath $q$}\,\rangle
\, = \,
- \frac{(2\,\pi)^3\,
  \delta^{(3)}(\mbox{\boldmath $p$}-\mbox{\boldmath $q$})\,m_e}
{\mbox{\boldmath $p$}^2-m_e\,E_0-i\,\epsilon} 
}
\nonumber\\[2mm] & &
- \frac{m_e}{\mbox{\boldmath $p$}^2-m\,E_0-i\,\epsilon}\,
  \frac{4\,\pi\,\alpha}
     {(\mbox{\boldmath $p$}-\mbox{\boldmath $q$})^2}\,
  \frac{m_e}{\mbox{\boldmath $q$}^2-m\,E_0-i\,\epsilon}
-R(\mbox{\boldmath $p$},\mbox{\boldmath $q$})
\,, \label{sumoverstates}
\end{eqnarray} 
where
\begin{eqnarray}
R(\mbox{\boldmath $p$},\mbox{\boldmath $q$}) & = &
\frac{64\,\pi\,\gamma^4}{\alpha\,(\mbox{\boldmath $p$}^2+\gamma^2)^2\,
       (\mbox{\boldmath $q$}^2+\gamma^2)^2}\,
\bigg[\,
\frac{5}{2} 
-\frac{4\,\gamma^2}{\mbox{\boldmath $p$}^2+\gamma^2}
-\frac{4\,\gamma^2}{\mbox{\boldmath $q$}^2+\gamma^2}
+\frac{1}{2}\,\ln A 
\nonumber\\[2mm] & &
\hspace{5cm}
+\frac{2\,A-1}{\sqrt{4\,A-1}}\arctan\Big(\sqrt{4\,A-1}\Big)
\,\bigg]
\,,
\label{Rpqfunction}
\\[2mm]
A & \equiv & 
\frac{(\mbox{\boldmath $p$}^2+\gamma^2)\,
      (\mbox{\boldmath $q$}^2+\gamma^2)}
     {4\,\gamma^2\,(\mbox{\boldmath $p$}-\mbox{\boldmath $q$})^2}
\,.
\end{eqnarray}
Details of the derivation of expression~(\ref{Rpqfunction}) can be
found in Ref.~\cite{Caswell2}. In Eq.~
(\ref{sumoverstates}),  
the three terms correspond to no Coulomb, one Coulomb, and two and more
Coulomb potentials in the intermediate state. In the bound state 
calculation of Eq.(\ref{Wfullformula}), the no-Coulomb contributions
are linearly divergent, the one-Coulomb contributions are logarithmically
divergent and the R-term contributions are finite.
In the case of the bound state contributions involving the R terms,
the following relation is quite useful ($q\equiv|\mbox{\boldmath
  $q$}|$)~\cite{Caswell2}:  
\begin{eqnarray}
\int\frac{d^3 \mbox{\boldmath $p$}}{(2\pi)^3}\,
R(\mbox{\boldmath $p$},\mbox{\boldmath $q$}) 
& = &
\frac{8\,\gamma^3}{\alpha\,(q^2+\gamma^2)^2}\,
\bigg(\,\frac{5}{2} - \ln 2 - 
   \frac{\gamma}{q}\,\arctan\Big(\frac{q}{\gamma}\Big)+
   \frac{1}{2}\,\ln\Big(1+\frac{q^2}{\gamma^2}\Big)-
   4\,\frac{\gamma^2}{q^2+\gamma^2}
\,\bigg)
\,.
\end{eqnarray}

The results for the
individual contributions of the RHS of Eq.~(\ref{Wfullformula}) read
\begin{eqnarray}
\lefteqn{
\langle\,1 {}^3S_1\,| \, V_4 \, |\, 1 {}^3S_1\,\rangle
\, = \,
\int\frac{d^3 \mbox{\boldmath $k_1$}}{(2\pi)^3}\,
\int\frac{d^3 \mbox{\boldmath $k_2$}}{(2\pi)^3}\,
\phi_0(\mbox{\boldmath $k_1$})\,
\bigg[\frac{2\,\pi\alpha}{m_e^2}\,d_1\bigg]
\phi_0(\mbox{\boldmath $k_2$})
\, = \,
\frac{m_e\,\alpha^4}{4}\,d_1
\,,
}
\label{WfullV4}
\\[4mm]
\lefteqn{
\langle\,1 {}^3S_1\,| \, V_{4\,\mbox{\tiny der}} \, 
      |\, 1 {}^3S_1\,\rangle
\, = \,
\int\frac{d^3 \mbox{\boldmath $k_1$}}{(2\pi)^3}\,
\int\frac{d^3 \mbox{\boldmath $k_2$}}{(2\pi)^3}\,
\phi_0(\mbox{\boldmath $k_1$})\,
\bigg[-\frac{4\,\pi\alpha}{3\,m_e^2}\,
  \frac{\mbox{\boldmath $k_1$}^2+\mbox{\boldmath $k_2$}^2}{m_e^2}
\bigg]
\phi_0(\mbox{\boldmath $k_2$})
}
\nonumber\\[2mm] & = &
-\frac{2\,m_e\,\alpha^5}{3\,\pi}\,\frac{\Lambda}{m_e} 
+\frac{m_e\,\alpha^6}{4}
\,,
\label{WfullV4der}
\\[4mm]
\lefteqn{
\langle\,1 {}^3S_1\,| \, V_4 \, 
  \sum\hspace{-7.5mm}\int\limits_{l\ne 1{}^3S_1}\,
  \frac{|\,l\,\rangle\,\langle \,l\,|}{E_0-E_l} \, V_4  
  \, |\, 1 {}^3S_1\,\rangle
}
\nonumber\\[2mm] & = &
\bigg[\,
-\frac{m_e \alpha^5}{4\,\pi}\,\frac{\Lambda}{m_e}
+\frac{m_e \alpha^6}{16}
\,\bigg]
-\bigg[\,
\frac{m_e \alpha^6}{8}\,\ln\Big(\frac{\Lambda}{2\,\gamma}\Big)
\,\bigg]
-\bigg[\,
\frac{3\,m_e \alpha^6}{16}
\,\bigg]
\,,
\label{WfullV4V4}
\\[4mm]
\lefteqn{
\bigg[\,
\langle\, 1 {}^3S_1\,| \,
V_4 \, \sum\hspace{-7.5mm}\int\limits_{l\ne 1{}^3S_1}\,
\frac{|\,l\,\rangle\,\langle \,l\,|}{E_0-E_l} \,
V_{\mbox{\tiny BF}}
\, |\, 1 {}^3S_1 \,\rangle + \mbox{h.c.}
\,\bigg]\,
}
\nonumber\\[2mm] & = &
-\bigg[\,
\frac{5\, m_e \alpha^5}{12\,\pi}\,\frac{\Lambda}{m_e}
+\frac{m_e \alpha^6}{4}\,
  \bigg( \frac{1}{12}-
    \ln\Big(\frac{\Lambda}{2\,\gamma}\Big)
  \bigg)
\,\bigg]
+\bigg[\,\frac{m_e \alpha^6}{8}\,
\bigg(
-1 + \frac{\pi^2}{3} - 
\frac{5}{3}\,\ln\Big(\frac{\Lambda}{2\,\gamma}\Big)
\bigg)
\,\bigg]
\nonumber\\[2mm] & &
+\bigg[\,\frac{m_e \alpha^6}{4}\,
\bigg(
1 - \frac{\pi^2}{6}
\bigg)
\,\bigg]
\,,
\label{WfullV4VBF}
\\[4mm]
\lefteqn{
\bigg[\,
\langle\, 1 {}^3S_1\,| \,
V_4 \, \sum\hspace{-7.5mm}\int\limits_{l\ne 1{}^3S_1}\,
\frac{|\,l\,\rangle\,\langle \,l\,|}{E_0-E_l} \,
\delta H_{\mbox{\tiny kin}}
\, |\, 1 {}^3S_1 \,\rangle + \mbox{H.c.}
\,\bigg]\,
}
\nonumber\\[2mm] & = &
\bigg[\,
\frac{m_e \alpha^5}{4\,\pi}\,\frac{\Lambda}{m_e}
-\frac{15\,m_e \alpha^6}{128}
\,\bigg]+
\bigg[\,
\frac{m_e \alpha^6}{8}\,\bigg(
-\frac{13}{32} + \ln\Big(\frac{\Lambda}{2\,\gamma}\Big)
\bigg)
\,\bigg]
+\bigg[\,
\frac{51\,m_e \alpha^6}{256}
\,\bigg]
\,,
\label{WfullV4Vkin}
\end{eqnarray}
where,  in the last three terms, the results from the no-Coulomb,
one-Coulomb and R-terms have been presented in separate brackets.
We note that for the bound state calculation we have adopted the usual
energy definition as given in Eq.~(\ref{NNLOSchroedinger}). 
A collection of integrals, which were useful to obtain the
results given above, can be found in Ref.~\cite{thesis}.

Adding all terms together and taking into account the corrections to
$d_1$ shown in Eq.~(\ref{d1final}), we arrive at the final result for
the single photon annihilation contributions to the hyperfine
splitting at NNLO~\cite{Hoang1}
\begin{eqnarray}
W^{\mbox{\tiny 1-$\gamma$ ann}} & = &
\frac{m_e \alpha^4}{4} 
-\frac{11\,m_e \alpha^5}{9\,\pi}
+ \frac{m_e \alpha^6}{4}\,\bigg[\,
\frac{1}{\pi^2}\,\bigg(\,
\frac{1477}{81} + \frac{13}{8}\,\zeta_3
\,\bigg)
- \frac{1183}{288} + \frac{9}{4}\,\ln 2 + 
\frac{1}{6}\,\ln\alpha^{-1}
\,\bigg]
\,.
\label{finalresult}
\end{eqnarray}
The $\ln\alpha^{-1}$ term in Eq.~(\ref{finalresult}) was already known
and is included in the $\ln\alpha^{-1}$ contribution quoted in
Eq.~(\ref{LOandNLO}). The single photon annihilation contribution to
the constant $K$ defined in Eq.~(\ref{LOandNLO}) corresponds to a
contribution of $-2.34$~MHz to the theoretical prediction of the
hyperfine splitting. The same result has been obtained in
Ref.~\cite{Adkins3} using the Bethe--Salpeter formalism and numerical
methods.

\vspace{1.5cm}
\section{Radial Excitations}
\label{sectionradial}
From the results presented in the previous sections it would be
straightforward to determine the single photon annihilation
contributions to the hyperfine splitting for arbitrary radial
excitations $n$: we ``only'' have to redo the bound state
calculation shown in the previous section
using the general $n{}^3S_1$ wave functions 
and the momentum space Coulomb Green function,
Eq.~(\ref{SchwingerGreenfunction}), where the $n{}^3S_1$ pole
is subtracted.
The matching calculation (i.e. the form of the short-distance
coefficient $d_1$), which does not depend on the non-relativistic
dynamics, would remain unchanged.  
Although such a strategy would be perfectly suited to the general
spirit of this work, it would be quite a cumbersome task to work our
all the formulae for arbitrary integer values of $n$. (Unlike the
contributions to the hyperfine splitting from the annihilation into
two~\cite{Adkins1} and three photons~\cite{Adkins2}, which are pure
short-distance corrections and therefore have a trivial dependence
on the value of $n$ 
($\propto |\phi_n(\mbox{\boldmath $0$})|^2\sim 1/n^3$), the single
photon annihilation 
contributions have a more complicated dependence on the value of $n$
because they involve a non-trivial mixing of bound state and
short-distance dynamics.) Thus for the calculation of the single
photon annihilation contributions 
to the hyperfine splitting for arbitrary values of $n$ we 
use a much simpler method, which one might almost call a ``back of the
envelope'' calculation~\cite{Hoang1}. However, we emphasize that this
method relies more on physical intuition and a careful inspection of
the ingredients needed for this specific calculation than on a
systematic approach that could be generally used for other problems as
well. Nevertheless, this method leads to the correct result, and we
therefore present it here as well, following Ref.~\cite{Hoang1}.     

We start from the formula for the single photon annihilation
contributions to the hyperfine splitting, Eq.~(\ref{Wfullformula}),
generalized to arbitrary values of $n$. (This amounts to replacing
``$1{}^3S_1$'' by ``$n{}^3S_1$'' everywhere in
Eq.~(\ref{Wfullformula}). In the following we simply refer to this
generalized equation as ``Eq.~(\ref{Wfullformula})''.) Because we use
Eq.~(\ref{Wfullformula}) only to identify two physical quantities
from which the single photon contributions to the hyperfine splitting
can be derived, we can consider the operators as unrenormalized
objects.
The relevant physical quantities are easily found if one
considers Eq.~(\ref{Wfullformula}) in configuration space
representation, where the operator $V_4$ corresponds to a
$\delta$-function. From this we see that Eq.~(\ref{Wfullformula})
depends entirely on the zero-distance Coulomb Green function 
$
A_n\equiv \langle\,\mbox{\boldmath $0$}\,| \,
\sum_{l\ne n}\hspace{-0.9cm}\int\hspace{0.5cm}\,
\frac{|\,l\,\rangle\,\langle \,l\,|}{E_l-E_n}
\, |\,\mbox{\boldmath $0$}\,\rangle
$
(where the $n{}^3\!S_1$ bound state pole is
subtracted) and on the rate for the annihilation of an $n{}^3\!S_1$ bound
state into a single photon,
$
P_n\equiv
\langle\,n\,| \, V_4 \, |\,n\,\rangle +
[
\langle\,n\,| \, V_4 \,\sum_{l\ne n}\hspace{-0.9cm}\int\hspace{0.5cm}\,
\frac{|\,l\,\rangle\,\langle \,l\,|}{E_n-E_l} \, 
(V_{\mbox{\tiny BF}} + \delta H_{\mbox{\tiny kin}})
\,|\,n\,\rangle + \mbox{h.c.}
] +
\langle\,n\,| \, V_{4\mbox{\tiny der}}
\, |\,n\,\rangle 
$
(where the effects from $V_{\mbox{\tiny BF}}$ and 
$\delta H_{\mbox{\tiny kin}}$ are included in the form of corrections
to the wave function). Because we have only considered unrenormalized
operators, $A_n$ and $P_n$ are still UV-divergent from the
integration over the high energy modes. In the NRQED approach worked
out in the previous sections the renormalization was achieved at the
level of the operators. Now, the renormalization will be carried
out by relating $A_n$ and $P_n$ to physical (and finite) quantities, which
incorporate the proper short-distance physics from the one photon
annihilation process. 
For $A_n$ this physical quantity is just the QED vacuum
polarization function in the non-relativistic limit and for $P_n$ the
Abelian contribution of the NNLO expression for the leptonic decay
width of a super-heavy quark--antiquark $n{}^3\!S_1$ bound
state~\cite{Bodwin1}. Both 
quantities have been determined in Refs.~\cite{Hoang3,Hoang4}. From
the results given in Refs.~\cite{Hoang3, Hoang4} it is 
straightforward to derive the renormalized versions of $A_n$ and
$P_n$, 
\begin{eqnarray}
A^{\mbox{\tiny phys}}_n & = &
\frac{m_e^2}{2\,\pi}\,\bigg\{\,
\frac{8}{9\,\pi} - \frac{\alpha}{2}\,\bigg[\,
C_1 + 
\bigg(\, \ln\Big(\frac{\alpha}{2\,n}\Big) - \frac{1}{n} + 
\gamma + \Psi(n)
\,\bigg)
\,\bigg]
\,\bigg\}
\,,
\\[2mm]
P^{\mbox{\tiny phys}}_n & = &
\frac{2\,\alpha\,\pi}{m_e^2}\,
\bigg(\frac{m_e^3\,\alpha^3}{8\,\pi\,n^3}\bigg)\,
\bigg\{\,
1 - 4\,\frac{\alpha}{\pi} + 
\alpha^2\,\bigg[\, C_2 - \frac{37}{24\,n^2} - 
\frac{2}{3}\,\bigg(\, \ln\Big(\frac{\alpha}{2\,n}\Big) - \frac{1}{n} + 
\gamma + \Psi(n)
\,\bigg)
\,\bigg]
\,\bigg\}
\,,
\label{renormalized}
\end{eqnarray}
where 
$
C_1  = 
\frac{1}{2\pi^2}(-3+\frac{21}{2}\zeta_3
) - \frac{11}{16} + \frac{3}{2}\ln 2
$
and
$
C_2  =  
\frac{1}{\pi^2}(\frac{527}{36}-\zeta_3) + 
\frac{4}{3}\ln 2 - \frac{43}{18}
$.
Here, $\gamma$ 
is the Euler constant and $\Psi$ the digamma
function.
Inserting now $A^{\mbox{\tiny phys}}_n$ and $P^{\mbox{\tiny phys}}_n$
back into expression~(\ref{Wfullformula}) we arrive at
\begin{equation}
W^{\mbox{\tiny 1-$\gamma$ ann}}_{n} \, = \, 
P^{\mbox{\tiny phys}}_n\,\bigg[\,
1- \frac{2\,\alpha\,\pi}{m_e^2}\,A^{\mbox{\tiny phys}}_n +
\bigg(\,\frac{2\,\alpha\,\pi}{m_e^2}\,A^{\mbox{\tiny phys}}_n\,\bigg)^2 
\,\bigg]
\,.
\label{finalsecond}
\end{equation}
which leads to~\cite{Hoang1} 
\begin{eqnarray}
W^{\mbox{\tiny 1-$\gamma$ ann}}_n & = &
\frac{m_e \alpha^4}{4\,n^3} 
-\frac{11\,m_e \alpha^5}{9\,\pi\,n^3}
\nonumber\\[2mm] & &
+ \frac{m_e \alpha^6}{4\,n^3}\,\bigg\{\,
\frac{1}{2}\,C_1 + C_2 + \frac{352}{81\,\pi^2} - \frac{37}{24\,n^2} 
+\frac{1}{6}\,\bigg[\, \frac{1}{n}
 + \ln\Big(\frac{2\,n}{\alpha}\Big) - \gamma  - \Psi(n)
\,\Bigg]
\,\bigg\}
\,.
\label{finalresultarbitraryn}
\end{eqnarray}
For the ground state $n=1$ Eq.~(\ref{finalresultarbitraryn}) reduces
to the result shown in
Eq.~(\ref{finalresult}). Equation~(\ref{finalresultarbitraryn}) has also
been confirmed by an independent calculation in
Ref.~\cite{Pachucki2}.

\vspace{1.5cm}
\section{Discussion}
\label{sectiondiscussion}

\par
\begin{table}[ht]
\begin{center}
\begin{tabular}{rllcrl} \hline
&       &               & Analytical/ &  & \\
& \raisebox{1.5ex}[-1.5ex]{Order} 
& \raisebox{1.5ex}[-1.5ex]{Specification} & numerical   
& \raisebox{1.5ex}[-1.5ex]{Contr. in MHz} 
& \raisebox{1.5ex}[-1.5ex]{Refs.} 
\\ \hline
1 & $m_e \alpha^4$ & & a & $204\,386.7\mbox{\hspace{7mm}}$ & \cite{Pirenne1} 
\\
2 & $m_e \alpha^5$ & & a & $-1\,005.5\mbox{\hspace{7mm}}$ & \cite{Karplus1}
\\
3 & $m_e \alpha^6 \ln\alpha^{-1}$ & & a & $19.1\mbox{\hspace{7mm}}$ &
    \cite{CaswellBodwin1} 
\\
4 & $m_e \alpha^6$ & non-recoil  & a & 
    $-10.43\mbox{\hspace{5mm}}$ &
    \cite{Sapirstein1,Pachucki2,Czarnecki1}+\cite{onephoton} 
\\
5 & $m_e \alpha^6$ & recoil (Caswell--Lepage) & n & 
    $3.1(6)\mbox{\hspace{2mm}}$ &
    \cite{Caswell1} 
\\
6 &  & recoil (Pachucki, Czarnecki {\it et al}) & a &
 $7.02\mbox{\hspace{5mm}}$ &
    \cite{Pachucki1,Czarnecki1}
\\
7 &  & recoil (Adkins--Sapirstein) & n & $1.32(7)$ &
    \cite{Adkins4}
\\
8 &  & 1-photon annihilation & a & $-2.34\mbox{\hspace{5mm}}$ &
    this work, \cite{Adkins3} 
\\
9 &  & 2-photon annihilation & a & $-0.61\mbox{\hspace{5mm}}$ &
    \cite{Adkins1}
\\
10 &  & 3-photon annihilation & a & $-0.97\mbox{\hspace{5mm}}$ &
    \cite{Adkins2}
\\
11 & $m_e \alpha^7 \ln^2\alpha^{-1}$ &  & a & $-0.92\mbox{\hspace{5mm}}$ &
    \cite{Karshenboim1,thesis}
\\ \hline
& & Sum (Caswell--Lepage) & & $203\,388.1(6)\mbox{\hspace{2mm}}$ &
\\ 
& & Sum (Pachucki, Czarnecki {\it et al})
  & & $203\,392.1\mbox{\hspace{7mm}}$ &
\\ 
& & Sum (Adkins--Sapirstein)  & & $203\,386.4\mbox{\hspace{7mm}}$ &
\\ 
& & Experiment  & & $203\,389.1(7)\mbox{\hspace{2mm}}$ &
  \cite{Ritter1}
\\
& &             & & $203\,387.5(1.6)\mbox{\hspace{-1mm}}$ &
  \cite{Mills1,Mills2}
\\ \hline
\end{tabular}
\parbox{15.6cm}{
\caption{\label{tab1}
Summary of theoretical calculations to the hyperfine splitting and
most the most recent experimental measurements.
} 
}
\end{center}
\end{table}
In Table~\ref{tab1} we have summarized the status of the theoretical
calculations to the positronium ground state hyperfine splitting,
including our own result. To order $m_e \alpha^6$ the contribution
that it logarithmic in $\alpha$ and the constant one are given
separately. The constant 
terms are further subdivided into non-recoil, recoil, and one-, two- and
three-photon annihilation contributions. The non-recoil corrections
correspond to diagrams in which one or two photons are emitted and
absorbed by the same lepton. (One example is the two-loop
contribution to the anomalous magnetic moment.) 
They are pure short-distance corrections and arise from loop momenta
of order $m_e$ and above.
In the effective field theory approach they are included as
a finite renormalization of the coefficients of the NRQED operators. 
 These non-recoil corrections were first
 evaluated numerically a certain time ago ~\cite{Sapirstein1}.
More recently, they were calculated    analytically by two
independent groups ~\cite{Pachucki2,Czarnecki1}
 who agreed with each other but disagreed
slightly with the numerical result (by about $0.10\,$ MHz).
The number we quote in the table is based on the
analytical expression.
The error of the result at order $m_e \alpha^4$ in row 1 is of the
level of a few $0.01\,$ MHz and not indicated explicitly. 
It comes from the
uncertainties in the Rydberg constant (Ref.~\cite{Rydberg}) and in 
$\alpha$ (Ref.~\cite{PDG}). The errors given in rows 5 and 7 are
of numerical origin. For all other contributions the errors are
negligible. The uncertainties due to the ignorance of the remaining 
$m_e \alpha^7 \ln\alpha^{-1}$ and $m_e \alpha^7$ contributions are
not taken into account in the summed results. 

Except for the recoil corrections, all the quoted results are by now  
well established. There is still some controversy concerning the
recoil corrections for which three different results can be found in
the literature. The first calculation was performed by Caswell and
Lepage in their seminal paper on NRQED~\cite{Caswell1}. Recently,
 new
 calculations   were performed by three different  groups,
 using  different techniques. First, 
   Pachucki~\cite{Pachucki1}, using  
an effective field theory approach in coordinate space and  a
different regularization scheme, obtained a result differing
significantly from the  one of Caswell and Lepage.
 Then  the Bethe--Salpeter formalism
was employed by Adkins and Sapirstein~\cite{Adkins4}, yielding 
yet another
 result. Finally, Czarnecki, Melnikov and
 Yelkhovsky~\cite{Czarnecki1}
performed the calculation in momentum space with dimensional
regularization. They obtained an analytical expression which agrees
with Pachucki's numerical result. The number quoted in the table
(row 6) corresponds to  their analytical expression.

Comparing with the most recent experimental measurement from
Ref.~\cite{Ritter1}, and ignoring remaining theoretical uncertainties,  
the result containing the Caswell--Lepage calculation for the recoil
contributions leads to an agreement between theory and experiment
($W_{\mbox{\tiny th}}-W_{\mbox{\tiny ex}}=-1.0(1.0)$~MHz), whereas the
prediction based on the result by Pachucki and Czarnecki {\it et al}
leads to a discrepancy of
more than four standard deviations 
($W_{\mbox{\tiny th}}-W_{\mbox{\tiny ex}}=3.0(0.7)$~MHz). On the other
hand, the Adkins--Sapirstein result differs by slightly 
less than four standard deviations but, in contradistinction with
Pachucki's result, it lies below the measured value:  
($W_{\mbox{\tiny th}}-W_{\mbox{\tiny ex}}=-2.7(0.7)$~MHz).
Recently, the NRQED calculation~\cite{Caswell1} has been repeated by 
one of us (P.L.), in collaboration with R. Hill. 
Preliminary results of this calculation are in contradiction with the
original NRQED calculation and also in agreement with Pachucki and
Czarnecki {\it et al}.
%
Hopefully, the theoretical situation will soon get settled. If
 the result of Pachucki
and Czarnecki {\it et al}   is indeed confirmed, the significant discrepancy
with experiment will have to be addressed. 

We note again that we have not included any estimates about the
remaining theoretical uncertainties in the considerations given above.
The next uncalculated corrections are of order $m_e \alpha^7 \,\ln
\alpha^{-1} \approx 0.7$~MHz and could significantly influence the
comparison between theory and experiment. However, the coefficient of
the log corrections are usually much smaller than 1, and we
therefore believe that a contribution of $1$~MHz from the higher 
order corrections is probably a conservative estimate. In this case,  
the discrepancy between theory and experiment remains
unexplained. Clearly, further work on positronium calculations is
necessary.

\vspace{1.5cm}
\section{Summary}
\label{sectionsummary}

We have provided the details of the  NRQED calculation 
of the ${\cal O} (m_e \alpha^6)$ contribution to the positronium 
ground state hyperfine splitting due to single photon annihilation
reported in an earlier paper. The counting rules needed to this order
have been explained in detail and a discussion on some of the
issues related to the use of an explicit cutoff on the momentum
integrals has been given. We have provided a list of integrals
useful for the evaluation of non-relativistic scattering diagrams.
Our result completes the ${\cal O}(m_e\alpha^6)$ calculation
of the ground state hyperfine splitting and permits a comparison
between theory and experiment at the level of 1 MHz.
A comparison with the most recent experimental measurement 
underlines the need for more theoretical work concerning the
${\cal O}(m_e\alpha^6)$ recoil corrections and higher order
contributions.   
\vspace{1.5cm}
\section*{Acknowledgements}
We are grateful to  A. Czarnecki,  P. Lepage, A.~V.~Manohar and
K. Melnikov for useful discussions and 
G.~Buchalla for reading the manuscript.
One of us (P.L.) has benefited from the hospitality of the Laboratory
of Nuclear Studies, Cornell University, where a part of this work has
been accomplished.
We thank T.~Teubner for providing us with Fig.~\ref{figrouting}.

\vspace{2cm}
\begin{appendix}
\section{Useful Integrals for the Matching Calculation}
\label{appendixshortdistance}
In this appendix we present a set of integrals that has been
useful for the matching calculations carried out in
Sec.~\ref{sectionmatching}. All integrals containing ultraviolet
divergences are regulated by the cutoff $\Lambda$, where the relation
$\Lambda\gg p_0$ is implied. Terms of order $1/\Lambda^k$, $k>0$ are
discarded. 
As explained in Sec.~\ref{sectionconcept} we have regulated all
infrared divergences by using a small fictitious photon mass
$\lambda$. In the following we give the results for arbitrary values
of $\lambda$, and in an expansion for $\lambda\ll p_0$ discarding
terms of order $\lambda^k$, $k>0$.  
For the matching calculations presented in this work only the fully
expanded results are relevant:
\begin{eqnarray}
{\cal{A}}_1 & = & 
\int\limits_0^{\Lambda\gg p_0} d p \,\frac{p^2}{p^2-p_0^2-i\epsilon}
\, = \,
\Lambda+i\,\pi\,\frac{p_0}{2}
\,,
\\[4mm]
{\cal{A}}_2 & = & 
\int\limits_0^{\Lambda\gg p_0} d p \,
\frac{p^2\,(p^2-p_0^2)}{(p^2+p_0^2+\lambda^2)^2-4\,p^2\,p_0^2}
\, = \,
\Lambda-\frac{3}{4}\,\lambda\,\pi 
\, \stackrel{\lambda\to 0}{\longrightarrow} \,
\Lambda
\,,
\\[4mm]
{\cal{A}}_3 & = & 
\int\limits_0^\infty d p \,\frac{p}{p^2-p_0^2-i\epsilon}\,
\ln\bigg(\frac{(p+p_0)^2+\lambda^2}{(p-p_0)^2+\lambda^2}\bigg)
\, = \,
\pi\,\arctan\Big(\frac{2\,p_0}{\lambda}\Big)+
 i\,\frac{\pi}{2}\,\ln\Big(1+\frac{4\,p_0^2}{\lambda^2}\Big)
\nonumber
\\[2mm]
& \stackrel{\lambda\to 0}{\longrightarrow} &
\frac{\pi^2}{2} + i\,\pi\,\ln\Big(\frac{2\,p_0}{\lambda}\Big)
\,,
\\[4mm]
{\cal{A}}_4 & = & 
\int\limits_0^{\Lambda\gg p_0} d p \,\frac{p^3}{p^2-p_0^2-i\epsilon}\,
\ln\bigg(\frac{(p+p_0)^2+\lambda^2}{(p-p_0)^2+\lambda^2}\bigg)
\nonumber
\\[2mm] & = &
p_0^2\,\pi\,\arctan\Big(\frac{2\,p_0}{\lambda}\Big) + 
p_0\,(4\,\Lambda-2\,\pi\,\lambda)+
i\,\frac{\pi\,p_0^2}{2}\,\ln\Big(1+\frac{4\,p_0^2}{\lambda^2}\Big)
\nonumber
\\[2mm]
& \stackrel{\lambda\to 0}{\longrightarrow} &
\frac{\pi^2}{2}\,p_0^2+4\,p_0\,\Lambda+
i\,\pi\,p_0^2\ln\Big(\frac{2\,p_0}{\lambda}\Big)
\,,
\\[4mm]
{\cal{A}}_5 & = & 
\int\limits_0^\infty d p \,
\frac{p}{(p^2+p_0^2+\lambda^2)^2-4\,p^2\,p_0^2}\,
\ln\bigg(\frac{(p+p_0)^2+\lambda^2}{(p-p_0)^2+\lambda^2}\bigg)
\nonumber
\\[2mm] & = &
\frac{\pi}{4\,\lambda\,p_0}\,\ln\Big(1+\frac{p_0^2}{\lambda^2}\Big)
\, \stackrel{\lambda\to 0}{\longrightarrow} \,
\frac{\pi}{2\,\lambda\,p_0}\,\ln\Big(\frac{p_0}{\lambda}\Big)
\,,
\\[4mm]
{\cal{A}}_6 & = & 
\int\limits_0^\infty d p \,
\frac{p^3}{(p^2+p_0^2+\lambda^2)^2-4\,p^2\,p_0^2}\,
\ln\bigg(\frac{(p+p_0)^2+\lambda^2}{(p-p_0)^2+\lambda^2}\bigg)
\nonumber
\\[2mm] & = &
\pi\,\arctan\Big(\frac{p_0}{\lambda}\Big)+
\pi\,\frac{p_0^2-\lambda^2}{4\,\lambda\,p_0}\,
 \ln\Big(1+\frac{p_0^2}{\lambda^2}\Big)
\, \stackrel{\lambda\to 0}{\longrightarrow} \,
\frac{\pi^2}{2}+
\frac{\pi\,p_0}{2\,\lambda}\,\ln\Big(\frac{p_0}{\lambda}\Big)
\,,
\\[4mm]
{\cal{A}}_7 & = & 
\int\limits_0^\infty d p \,
\frac{p}{p^2-p_0^2-i\, \epsilon}\,
\frac{1}{(p^2+p_0^2+\lambda^2)^2-4\,p^2\,p_0^2}\,
\ln\bigg(\frac{(p+p_0)^2+\lambda^2}{(p-p_0)^2+\lambda^2}\bigg)
\nonumber
\\[2mm] & = &
\frac{\pi}{4\,\lambda^2\,p_0\,(\lambda^2+4\,p_0^2)}\,
\bigg[\, 
  4\,p_0\,\bigg(\,\arctan\Big(\frac{2\,p_0}{\lambda}\Big)-
             \arctan\Big(\frac{p_0}{\lambda}\Big)\,\bigg)-
  \lambda\,\ln\Big(1+\frac{p_0^2}{\lambda^2}\Big)
\,\bigg]
\nonumber
\\[2mm] & & \hspace{2cm} +
i\,\frac{\pi}{2\,\lambda^2\,(\lambda^2+4\,p_0^2)}\,
\ln\Big(1+\frac{4\,p_0^2}{\lambda^2}\Big)
\nonumber
\\[2mm]
& \stackrel{\lambda\to 0}{\longrightarrow} &
\pi\,\bigg[\,
\frac{4\,p_0+i\,\lambda}{32\,\lambda\,p_0^4} - 
\frac{1}{8\,\lambda\,p_0^3}\,\ln\Big(\frac{p_0}{\lambda}\Big) -
i\,\frac{\lambda^2-4\,p_0^2}{16\,\lambda^2\,p_0^4}\,
  \ln\Big(\frac{2\,p_0}{\lambda}\Big)
\,\bigg]
\,,
\\[4mm]
{\cal{A}}_8 & = & 
\int\limits_0^\infty d p_1 \,
\int\limits_0^{\Lambda\gg p_0} d p_2 \,
\frac{p_1\,p_2}{(p_1^2-p_0^2-i\epsilon)\,(p_2^2-p_0^2-i\epsilon)}\,
\ln\bigg(\frac{(p_1+p_2)^2+\lambda^2}{(p_1-p_2)^2+\lambda^2}\bigg)
\nonumber
\\[2mm] & = &
\pi^2\,\bigg[\,
\ln\Big(\frac{\Lambda}{\lambda}\Big)-
  \frac{1}{2}\,\ln\Big(1+\frac{4\,p_0^2}{\lambda^2}\Big)
  + i\,\arctan\Big(\frac{2\,p_0}{\lambda}\Big)
\,\bigg]
\nonumber
\\[2mm]
& \stackrel{\lambda\to 0}{\longrightarrow} &
\pi^2\,\bigg[\, 
  -\ln\Big(\frac{2\,p_0}{\Lambda}\Big)+i\,\frac{\pi}{2}
\,\bigg]
\,,
\\[4mm]
{\cal{A}}_9 & = & 
\int\limits_0^\infty d p_1 \,
\int\limits_0^{\Lambda\gg p_0} d p_2 \,
\frac{p_1}{p_1^2-p_0^2-i\epsilon}\,
\frac{p_2\,(p_2^2-p_0^2)}{(p_2^2+p_0^2+\lambda^2)^2-4\,p_2^2\,p_0^2}\,
\ln\bigg(\frac{(p_1+p_2)^2+\lambda^2}{(p_1-p_2)^2+\lambda^2}\bigg)
\nonumber
\\[2mm] & = &
-\pi^2\,\bigg[\,
\frac{\lambda}{4\,p_0}\,\arctan\Big(\frac{p_0}{\lambda}\Big)+\ln 2 +
\frac{1}{4}\,\ln\Big(1+\frac{p_0^2}{\lambda^2}\Big)-
\ln\Big(\frac{\Lambda}{\lambda}\Big)
\,\bigg] +
\nonumber
\\[2mm] & & \hspace{2cm}
i\,\pi^2\,\bigg[\, 
\frac{1}{2}\,\arctan\Big(\frac{p_0}{\lambda}\Big)-
\frac{\lambda}{8\,p_0}\,\ln\Big(1+\frac{p_0^2}{\lambda^2}\Big)
\,\bigg]
\nonumber
\\[2mm]
& \stackrel{\lambda\to 0}{\longrightarrow} &
\pi^2\,\bigg(\,\ln\Big(\frac{\Lambda}{\lambda}\Big)-
\frac{1}{2}\,\ln\Big(\frac{p_0}{\lambda}\Big)-
\ln 2\,\bigg)+i\,\frac{\pi^3}{4}
\,,
\\[4mm]
{\cal{A}}_{10} & = & 
\int\limits_0^\infty d p_1 \,
\int\limits_0^\infty d p_2 \,
\frac{p_1}{(p_1^2-p_0^2-i\epsilon)\,(p_2^2-p_0^2-i\epsilon)}\,
\ln\bigg(\frac{(p_1+p_2)^2+\lambda^2}{(p_1-p_2)^2+\lambda^2}\bigg)\,
\ln\bigg(\frac{(p_2+p_0)^2+\lambda^2}{(p_2-p_0)^2+\lambda^2}\bigg)
\nonumber
\\[2mm] & = &
\frac{\pi^2}{p_0}\,\bigg[\,
-\frac{\pi^2}{6}+
\frac{1}{2}\,\arctan^2\Big(\frac{2\,p_0}{\lambda}\Big)-
\frac{3}{8}\,\ln^2\Big(1+\frac{4\,p_0^2}{\lambda^2}\Big)-
{\rm Li}_2\Big(-1-\frac{2\,i\,p_0}{\lambda}\Big)
\nonumber
\\[2mm] &  & \hspace{1cm} 
- {\rm Li}_2\Big(-1+\frac{2\,i\,p_0}{\lambda}\Big)
\,\bigg]
\nonumber
\\[2mm] & &
+i\,\frac{\pi^2}{p_0}\,\bigg[\,
 \arctan\Big(\frac{2\,p_0}{\lambda}\Big)\,\bigg( 
   \ln 2+\ln\Big(1+\frac{4\,p_0^2}{\lambda^2}\Big)  \bigg)
\nonumber
\\[2mm] & & \hspace{0.5cm}
+i\,\bigg(
{\rm Li}_2\Big(-1+\frac{2\,i\,p_0}{\lambda}\Big)-
{\rm Li}_2\Big(-1-\frac{2\,i\,p_0}{\lambda}\Big)+
\frac{1}{2}\,{\rm Li}_2\Big(\frac{1}{2}-\frac{i\,p_0}{\lambda}\Big)-
\frac{1}{2}\,{\rm Li}_2\Big(\frac{1}{2}+\frac{i\,p_0}{\lambda}\Big)
\bigg)
\,\bigg]
\nonumber
\\[2mm]
& \stackrel{\lambda\to 0}{\longrightarrow} &
\frac{\pi^2}{2\,p_0}\,\bigg[\,
\frac{\pi^2}{12}-\ln^2\Big(\frac{2\,p_0}{\lambda}\Big)
+i\,\pi\ln\Big(\frac{2\,p_0}{\lambda}\Big)
\,\bigg]
\,,
\\[4mm]
{\cal{A}}_{11} & = & 
\int\limits_0^{\Lambda\gg p_0} d p_1 \,
\int\limits_0^\infty d p_2 \,
\frac{p_1}{p_2^2-p_0^2-i\epsilon}\,
\ln\bigg(\frac{(p_1+p_2)^2+\lambda^2}{(p_1-p_2)^2+\lambda^2}\bigg)\,
\ln\bigg(\frac{(p_2+p_0)^2+\lambda^2}{(p_2-p_0)^2+\lambda^2}\bigg)
\nonumber
\\[2mm] & = &
4\,\pi\,\Lambda\,\bigg[\,
\arctan\Big(\frac{2\,p_0}{\lambda}\Big)+
 \frac{i}{2}\,\ln\Big(1+\frac{4\,p_0^2}{\lambda^2}\Big)
\,\bigg] -
2\,\pi^2\,p_0 -
2\,\lambda\,\pi^2\,\arctan\Big(\frac{2\,p_0}{\lambda}\Big)
\nonumber
\\[2mm] &  & \hspace{1cm}
- i\,\lambda\,\pi^2\,\ln\Big(1+\frac{4\,p_0^2}{\lambda^2}\Big)
\nonumber
\\[2mm]
& \stackrel{\lambda\to 0}{\longrightarrow} &
2\,\Lambda\,\pi^2 - 2\,p_0\,\pi^2 +
4\,i\,\Lambda\,\pi\,\ln\Big(\frac{2\,p_0}{\lambda}\Big)
\,,
\\[4mm]
{\cal{A}}_{12} & = & 
\int\limits_0^\infty d p_1 \,
\int\limits_0^{\Lambda\gg p_0} d p_2 \,
\frac{p_1}{p_1^2-p_0^2-i\epsilon}\,
\ln\bigg(\frac{(p_1+p_2)^2+\lambda^2}{(p_1-p_2)^2+\lambda^2}\bigg)\,
\ln\bigg(\frac{(p_2+p_0)^2+\lambda^2}{(p_2-p_0)^2+\lambda^2}\bigg)
\nonumber
\\[2mm] & = &
\pi^2\,\bigg[\,
-4\,\lambda\,\arctan\Big(\frac{p_0}{\lambda}\Big)+
2\,p_0\,\bigg(\,2-2\,\ln 2-
  \ln\Big(\frac{\lambda^2+p_0^2}{\Lambda^2}\Big)\,\bigg)
\,\bigg]
\nonumber
\\[2mm] & & \hspace{0.5cm}
i\,\pi^2\,\bigg[\,
  4\,p_0\,\arctan\Big(\frac{p_0}{\lambda}\Big)-
  2\,\lambda\,\ln\Big(1+\frac{p_0^2}{\lambda^2}\Big)
\,\bigg]
\nonumber
\\[2mm]
& \stackrel{\lambda\to 0}{\longrightarrow} &
4\,p_0\,\pi^2\,\bigg[\,
  1-\ln\Big(\frac{2\,p_0}{\Lambda}\Big)+i\,\frac{\pi}{2}
\,\bigg]
\,.
\end{eqnarray}

\end{appendix}

\vspace{1.5cm}
%
%
%
\sloppy
\raggedright
\def\app#1#2#3{{\it Act. Phys. Pol. }{\bf B #1} (#2) #3}
\def\apa#1#2#3{{\it Act. Phys. Austr.}{\bf #1} (#2) #3}
\def\lhc{Proc. LHC Workshop, CERN 90-10}
\def\npb#1#2#3{{\it Nucl. Phys. }{\bf B #1} (#2) #3}
\def\nP#1#2#3{{\it Nucl. Phys. }{\bf #1} (#2) #3}
\def\plb#1#2#3{{\it Phys. Lett. }{\bf B #1} (#2) #3}
\def\prd#1#2#3{{\it Phys. Rev. }{\bf D #1} (#2) #3}
\def\pra#1#2#3{{\it Phys. Rev. }{\bf A #1} (#2) #3}
\def\pR#1#2#3{{\it Phys. Rev. }{\bf #1} (#2) #3}
\def\prl#1#2#3{{\it Phys. Rev. Lett. }{\bf #1} (#2) #3}
\def\prc#1#2#3{{\it Phys. Reports }{\bf #1} (#2) #3}
\def\cpc#1#2#3{{\it Comp. Phys. Commun. }{\bf #1} (#2) #3}
\def\nim#1#2#3{{\it Nucl. Inst. Meth. }{\bf #1} (#2) #3}
\def\pr#1#2#3{{\it Phys. Reports }{\bf #1} (#2) #3}
\def\sovnp#1#2#3{{\it Sov. J. Nucl. Phys. }{\bf #1} (#2) #3}
\def\sovpJ#1#2#3{{\it Sov. Phys. JETP Lett. }{\bf #1} (#2) #3}
\def\jl#1#2#3{{\it JETP Lett. }{\bf #1} (#2) #3}
\def\jet#1#2#3{{\it JETP }{\bf #1} (#2) #3}
\def\zpc#1#2#3{{\it Z. Phys. }{\bf C #1} (#2) #3}
\def\ptp#1#2#3{{\it Prog.~Theor.~Phys.~}{\bf #1} (#2) #3}
\def\nca#1#2#3{{\it Nuovo~Cim.~}{\bf #1A} (#2) #3}
\def\ap#1#2#3{{\it Ann. Phys. }{\bf #1} (#2) #3}
\def\hpa#1#2#3{{\it Helv. Phys. Acta }{\bf #1} (#2) #3}
\def\ijmpA#1#2#3{{\it Int. J. Mod. Phys. }{\bf A #1} (#2) #3}
\def\ZETF#1#2#3{{\it Zh. Eksp. Teor. Fiz. }{\bf #1} (#2) #3}
\def\jmp#1#2#3{{\it J. Math. Phys. }{\bf #1} (#2) #3}
\def\yf#1#2#3{{\it Yad. Fiz. }{\bf #1} (#2) #3}
\def\epjc#1#2#3{{\it Eur. Phys. J. C }{\bf #1} (#2) #3}

%
%
\end{document}